\documentclass[preprintnumbers,amsmath,amssymb,prd,superscriptaddress,notitlepage,longbibliography]{revtex4-1}
\pdfoutput=1

\usepackage{graphicx}
\usepackage{dcolumn}
\usepackage{bm}
\usepackage{wasysym}
\usepackage{threeparttable}

\newcommand{\frachalf}{\frac{1}{2}}

\newcommand{\be}{\begin{equation}}
\newcommand{\ee}{\end{equation}}

\newcommand{\sN}{\mathcal{N}}

\newcommand{\vtheta}{\mathbf{\theta}}

\newcommand{\vb}{\mathbf{b}}

\newcommand{\vc}{\mathbf{c}}

\newcommand{\vmu}{\mathbf{\mu}}

\newcommand{\vphi}{\mathbf{\phi}}

\newcommand{\vep}{\mathbf{\epsilon}}
\newcommand{\vo}{\mathbf{o}}

\newcommand{\vS}{\mathbf{S}}
\newcommand{\vC}{\mathbf{C}}

\newcommand{\vQ}{\mathbf{Q}}

\newcommand{\vP}{\mathbf{P}}
\newcommand{\vN}{\mathbf{N}}
\newcommand{\vA}{\mathbf{A}}

\newcommand{\vK}{\mathbf{K}}

\newcommand{\vR}{\mathbf{R}}

\newcommand{\vW}{\mathbf{W}}

\newcommand{\vI}{\mathbf{I}}

\newcommand{\vD}{\mathbf{D}}

\newcommand{\mnras}{{\em Mon. Not. Roy. Astron. Soc. }}

\newcommand{\jcap}{JCAP}
\newcommand{\physrep}{{\em Phys. Rept. }}
\newcommand{\aap}{{\em Astron. Astrophys. }}

\def\prd{{\em Phys. Rev. }{\bf D }}

\def\pvmhid#1{}

\def\afhid#1{}

\def\ashid#1{}

\newcommand{\vQp}{\vQ^\prime}

\begin{document}

\title{Renormalization group computation of likelihood functions for 
cosmological data sets}

\author{Patrick McDonald}
\email{PVMcDonald@lbl.gov}
\affiliation{Lawrence Berkeley National Laboratory, One Cyclotron Road,
Berkeley, CA 94720, USA}

\date{\today}

\begin{abstract}

I show how a renormalization group (RG) method can be used to incrementally 
integrate the information in cosmological large-scale structure data sets 
(including CMB, galaxy redshift surveys, etc.). I show numerical tests for 
Gaussian fields, where the method allows arbitrarily close to exact computation
of the likelihood function in order $\sim N$ time, even for problems with no 
symmetry, compared to $N^3$ for brute force linear algebra (where $N$ is the 
number of data points -- to be fair, methods already exist to solve the 
Gaussian problem in at worst $N \log N$ time, and this method will not 
necessarily be faster in practice).  The method
requires no sampling or other Monte Carlo (random) element.  
Non-linearity/non-Gaussianity can be accounted for to the extent that terms 
generated by integrating out small scale modes can be projected onto a 
sufficient basis, e.g., at least in the sufficiently perturbative regime. The 
formulas to evaluate are straightforward and require no understanding of 
quantum field theory, but this paper may also serve as a pedagogical 
introduction to Wilsonian RG for astronomers.

\end{abstract}

\maketitle

\section{Introduction}

Structure in the Universe starts as a near perfect Gaussian random field, and
remains tightly connected to the initial conditions on large scales
\cite{1993ppc..book.....P}, 
motivating using a rigorous likelihood analysis to extract information about
models from observed large scale fluctuations. 
Unfortunately, the likelihood function one can write 
\cite[e.g.,][]{1998PhRvD..57.2117B}
is generally 
cumbersome to evaluate numerically for realistic large datasets,
with time to compute naively scaling like 
$N^3$ using brute force linear algebra (where $N$ is the number of data 
points), which quickly becomes hopeless.
Various methods have been devised to speed this up to at worst 
$N \log N$ 
\cite[e.g.,][]{2003MNRAS.346..619P,2003NewA....8..581P,2004PhRvD..70h3511W,
2007PhRvD..76d3510S,2008MNRAS.389..497K,
2017JCAP...12..009S,2018JCAP...01..003F}, 
but in practice well-motivated but not explicitly likelihood-based 
power/correlation estimation methods are usually used instead
\citep[e.g.,][]{2016A&A...594A..11P,2017MNRAS.464.3409B}. 
Here I describe what seems like a particularly elegant way to make 
likelihood computation fast enough to be practical, 
based on renormalization group (RG) ideas from
quantum field theory \cite{1974PhR....12...75W,2008mqft.book.....B}. 

To be clear, the purpose of this paper is not to argue that this is the 
best/fastest way to estimate power spectra of Gaussian fields. For example,
many will 
inevitably consider the approach of defining semi-heuristic 
``summary statistics,'' like any
reasonably computed power spectrum or correlation function of the data, 
backed up by 
calibration and error estimation using mock data, to be more appealing to code,
surely faster, and optimal enough for their purposes. 
No doubt this is true a lot of the time. 
The point here is to present a theoretical physics-oriented framework for 
viewing the LSS analysis problem, which may present new possibilities for 
extension to, e.g., the largest scales where pseudo-$C_\ell$-type methods are
least applicable, and a new way to approach non-linearities. The spirit of 
this introductory paper is summarized by the old physics saying that
``any good theorist can tell you several different 
ways to solve the same problem.'' 

While it isn't at all difficult once you get into it, 
the math here may look a little bit technical at first glance. 
The basic idea is 
straightforward though: 
First, we effectively integrate over the highest $k$ Fourier 
modes, i.e., smallest scale structure in the problem,
absorbing the numbers that result from this integration 
into redefinitions of constants in the likelihood function.
(For an extreme pedagogical example, suppose we were asked to do the integral 
$I=A \int dx \int dy~ f(x)g(y)$,
and we knew how to do the integral over $y$ to produce a number 
$I_y\equiv \int dy~ g(y)$. Obviously
we can do the integral over $y$, 
define $A^\prime \equiv A I_y$, and now write the original integral as
$I=A^\prime \int dx ~f(x)$. Trivial as it seems, this is basically
renormalization in a nutshell.) Operationally,
once the smallest scale structure is gone (already accounted for), 
we can ``coarse grain'', i.e., combine small cells in the data set
into bigger cells. This coarse graining is critical because we can't 
integrate over all Fourier
modes at once -- that takes us back to the impossible problem we are trying to 
get around. 
After coarse graining, we can integrate out the smallest remaining
structure again, 
then coarse grain further, and so on. It may be puzzling that you 
will not see any explicit integration over modes in the paper, 
even though that is 
presented as the key idea in this kind of RG procedure.
Explicit integration over modes is
generally replaced by suppression of their fluctuations for technical reasons
\cite{1974PhR....12...75W,2008mqft.book.....B}, but with enough thought one
can see that these things are effectively equivalent. 

\section{Effectively exact Gaussian likelihood evaluation}

We will primarily discuss Gaussian likelihood functions, where it is easy
to make the calculations effectively
exact, although the basic RG equation we derive will be 
exact in principle even for
non-Gaussian likelihood functions (with non-Gaussianity, 
approximations are usually needed to evaluate it). 

Suppose we have an observed 
data vector $\vo$, linearly 
related to an underlying Gaussian field of interest, 
$\vphi$, by $\vo=\vmu+\vR\vphi +\vep$, where $\vmu$ is the mean vector, 
$\vR$ is a short-range response  
matrix and $\vep$ is Gaussian error with
covariance matrix $\vN$, assumed to contain at most short-range 
correlation. Assume 
$\vo$ has $N_o$ elements and $\vphi$ has $N_\phi$, i.e., 
$\vR$ is an $N_o \times N_\phi$ matrix, where fundamentally we should take
$N_\phi\rightarrow \infty$ because $\phi$ is a continuous field 
(in practice in this paper I leave $N_\phi=N_o$ to start the computation --
from there a key point of the RG method is to reduce this by coarsenings).
For now assume $\vmu$ is known and subtract it
from $\vo$ (the mean or its parameters can be fit for using 
the method we discuss). 
The covariance matrix (power spectrum in Fourier basis) of $\vphi$ 
is $\vP(\vtheta)$, which depends on parameters $\vtheta$ 
which we are ultimately trying to measure. 
(In the calculations here I will assume the signal field can be treated as 
translation invariant so $\vP$ is truly diagonal in Fourier space, however, 
this is not strictly true for the large-scale evolving Universe. 
After working through the idealized case, it will become clear how to 
generalize to a less idealized real Universe.)  
Dropping irrelevant Bayesian factors relating 
$L(\vo|\vphi)L(\vphi|\vtheta)$ to $L(\vphi,\vtheta| \vo)$,
the likelihood function for $\vtheta$ and $\vphi$ given $\vo$ is
\be
L(\vphi,\vtheta| \vo)= \frac{
e^{-\frac{1}{2}\vphi^t \vP^{-1}\vphi -
\frac{1}{2}(\vo-\vR\vphi)^t \vN^{-1}(\vo-\vR\vphi)} }{
\sqrt{\det(2\pi \vP)
\det(2\pi \vN)}}~.
\ee

We are really only interested in $L(\vtheta|\vo)$, so the standard approach
is to perform the Gaussian integral over $\vphi$ to obtain:
\be
L(\vtheta| \vo)=\int d\vphi ~L(\vphi,\vtheta| \vo)= \frac{
e^{-\frac{1}{2}\vo^t \vC^{-1}\vo}}
{\sqrt{\det(2 \pi \vC)}}
\ee
where $\vC=\vR \vP\vR^t+\vN$. 
From here we can, e.g., take $\vtheta$ derivatives to define
the standard band power estimator \cite[e.g.,][]{1998PhRvD..57.2117B}. 
For large data sets, however, the problem of 
numerically evaluating the 
resulting equations is usually considered too difficult to bother doing
(to make a long story short).

The key to the difficulty in the standard approach is that $\vP$ is diagonal 
in Fourier space but dense in real space, while $\vN$ generally vice versa. 
This makes $\vC$ dense in both. Computing the inverse or determinant of 
$\vC$ scales like $N^3$, making it impossible to do using brute force linear
algebra routines for large
data sets.  
The approach of this paper systematically solves this problem.
(Alternative solutions include conjugate gradient $\vC^{-1}\vo$ multiplication 
combined
with a stochastic determinant calculation, and the approach of 
\cite{2017JCAP...12..009S}.) 

\subsection{General RG equation}

Consider the integral 
\be
I\equiv \int d\vphi~ e^{-S(\vphi)}\equiv 
\int d\vphi~ e^{-\frachalf \vphi^t \vQ^{-1} \vphi
- S_I(\vphi)-\sN}~.
\label{eq:IderiveRG}
\ee
If we take
$\vQ^{-1}=\vP^{-1}$,
$S_I(\phi)= 
\frachalf \vphi^t \vR^t\vN^{-1}\vR \vphi-\vo^t \vN^{-1}\vR \vphi$, and
$\sN=\frachalf {\rm Tr}\ln(2\pi \vP)+\frachalf {\rm Tr}\ln(2\pi \vN)
+\frac{1}{2}\vo^t \vN^{-1}\vo$, this integral is 
exactly $L(\vtheta|\vo)$.
The $\ln \vP$ term represents the normalizing factor $\det \vP$, which
we need to keep around more carefully than usual because it 
depends on $\vtheta$. The $\vN$ parts of $\sN$ could usually be dropped, 
assuming $\vN$ does not depend on parameters. 
If we generalize to 
$\vQ^{-1}=\vP^{-1}\vW^{-1}+\vK$, 
then $I$ is the 
$L(\vtheta|\vo)$ that we want if $\vW\rightarrow 1$ and 
$\vK\rightarrow 0$.
On the other hand,
if $\vK\rightarrow \infty$, this term completely dominates the integral
making it trivial to evaluate, as long as
$\vK$ is chosen to be diagonal in Fourier basis. 
The integration could 
alternatively be cut off by $\vW\rightarrow 0$, or we could use
$\vK$ together with $\vW$ to modify the detailed behavior of the cutoff, but
in this paper we will mostly assume $\vW=1$ (because keeping
the cutoff term distinct from $\vP$ initially seemed simpler). 
Our RG strategy will be to dial $\vK$ from 0 to 
infinity, while redefining constants in the likelihood function to 
guarantee that the
result of the integral does not change.
We will parameterize this dialing 
by a length scale $\lambda$, with 
$\vK(k\lambda<<1) \rightarrow 0$ and
$\vK(k\lambda>>1) \rightarrow \infty$, i.e., for any given $\lambda$
smaller scale modes will already be suppressed (i.e., driven to zero 
variance) by $\vK\rightarrow \infty$,  
while larger scale modes will be untouched with
$\vK\rightarrow 0$. 
$\lambda$ will start at 0, i.e., 
$\vK= 0$ for all modes, and go toward $\infty$, i.e.,
$\vK= \infty$ for all modes.  Generally we can choose the 
details of $\vK(k,\lambda)$ in any way that looks useful. For the Gaussian
problem,
the constants we will redefine to preserve the integral value as $\vQ$
changes are the quadratic coefficient of $S_I$, which we will call
$\vA(\lambda)$, with $\vA(\lambda=0)=\vR^t\vN^{-1}\vR$, the linear 
coefficient of $S_I$, which we will call $\vb(\lambda)$, with 
$\vb(\lambda=0)=\vR^t\vN^{-1}\vo$, and the constant $\sN(\lambda)$, 
with 
$\sN(\lambda=0)=\frachalf {\rm Tr}\ln(2\pi \vP)+
\frachalf {\rm Tr}\ln(2\pi \vN)+\frac{1}{2}\vo^t \vN^{-1}\vo$. 
In other words, for the Gaussian problem the 
renormalized $S_I$ will be $S_I(\lambda)=
\frachalf \vphi^t \vA(\lambda) \vphi-\vb^t(\lambda)\vphi$. Note, however,
that for the rest of this subsection the calculation will apply for 
general $S_I(\phi)$, i.e., not just quadratic in $\phi$.

To preserve the value of the integral as $\lambda$ changes, we require the
derivative with respect to $\lambda$ to be zero, i.e., 
\be
\frac{\partial I}
{\partial \ln \lambda}=
-\int d\vphi~ 
\left(\frachalf \vphi^t \frac{\partial \vQ^{-1} }
{\partial \ln \lambda} \vphi + 
\frac{\partial S_I}{\partial \ln \lambda}
+\frac{\partial \sN}{\partial \ln \lambda}
\right)
e^{-S(\vphi)} = 0~.
\label{eq:dIdt}
\ee 
The most obvious way to change $S_I$ and $\sN$
to compensate the change in $\vQ$ would be to 
adjust the quadratic term in $S_I$ ($\vA$ in the definition above) to leave 
the net quadratic term 
in the integral unchanged, but this would 
literally accomplish nothing.   
To find the alternative changes that will effect the integration over 
$\phi$ we follow most closely \cite{2008mqft.book.....B}. 
Note that
\be
\frachalf \vphi^t \frac{\partial \vQ^{-1}}
{\partial \ln \lambda} \vphi
~e^{-\frachalf \vphi^t \vQ^{-1} \vphi}  
=
-\frachalf \vphi^t \vQ^{-1} \vQp \vQ^{-1}\vphi
~e^{-\frachalf \vphi^t \vQ^{-1} \vphi}  
=
-\frachalf\left[ \partial_\vphi^t \vQp
\partial_\vphi 
+{\rm Tr}\left(\vQ^{-1} \vQp \right)
\right]
~e^{-\frachalf \vphi^t \vQ^{-1} \vphi}  
\ee
where we have defined $^\prime$ to mean $\partial_{\ln \lambda}$,
so we have
\begin{eqnarray}
\int &d\vphi&~ 
\frachalf \vphi^t \frac{\partial \vQ^{-1}}
{\partial \ln \lambda} \vphi 
~e^{-\frachalf \vphi^t \vQ^{-1} \vphi - S_I(\vphi)-\sN} 
=
-\frachalf
\int d\vphi~  e^{- S_I(\vphi)-\sN}\left[
\partial_\vphi^t \vQp \partial_\vphi 
+{\rm Tr}\left(\vQ^{-1} \vQp \right) \right]
e^{-\frachalf \vphi^t \vQ^{-1} \vphi}  \\ \nonumber
&=& 
-\frachalf
\int d\vphi~  
e^{-\frachalf \vphi^t \vQ^{-1} \vphi-\sN}  
\left[
\partial_\vphi^t \vQp \partial_\vphi 
+{\rm Tr}\left(\vQ^{-1} \vQp\right)
\right]
e^{- S_I(\vphi)} \\ \nonumber
&=&\frachalf
\int d\vphi~ 
\left( 
{\rm Tr}\left[ \vQp \left(
\frac{\partial^2 S_I}
{\partial \vphi \partial \vphi^t}-\vQ^{-1}\right)
\right]
-
\frac{\partial S_I}
{\partial \vphi^t}
\vQp
\frac{\partial S_I}
{\partial \vphi}
\right)
e^{-S(\vphi)}
\end{eqnarray}
where we integrated by parts to make the $\vphi$ derivatives
act on $e^{- S_I(\vphi)}$. 
To satisfy Eq. (\ref{eq:dIdt}) we need:
\be
S_I^\prime
+\sN^\prime=
\frachalf
\frac{\partial S_I}
{\partial \vphi^t}
\vQp
\frac{\partial S_I}
{\partial \vphi}
+\frachalf {\rm Tr}\left[
\vQp
\left(
\vQ^{-1}-
\frac{\partial^2 S_I}
{\partial \vphi \partial \vphi^t}\right)
\right]
\label{eq:masterRG}
\ee
If we dropped the $\phi$-independent pieces, which are normalization factors 
that are irrelevant if we are only going to compute expectation values with
$e^{-S}$ weight, this is a completely standard RG equation
\cite[e.g.,][]{2008mqft.book.....B}. 
For now we do not want to give up on computing exact likelihoods as a function
of parameters, so we keep the normalization factors.
Note that the derivation is
completely general in the sense that we made no assumption about 
the details of $S_I$ and $\vQ$ (e.g., $S_I$ can be an arbitrary, not 
necessarily quadratic function of $\phi$).

\subsection{RG equations for the Gaussian problem \label{sec:GaussianRG}}

The interesting content comes from
expanding the derivatives of $S_I$.
As mentioned above, we define for the Gaussian
problem
$S_I= \frachalf \vphi^t \vA \vphi-\vb^t\vphi$.
Plugging this into Eq. (\ref{eq:masterRG}) gives
\be
\frachalf \vphi^t\vA^\prime\vphi -\vb^{\prime t} \vphi 
+\sN^\prime=
\frachalf
\left(\vA \vphi-\vb\right)^t
\vQp
\left(\vA \vphi-\vb\right)
+\frachalf {\rm Tr}\left[
\vQp
\left(\vQ^{-1}-\vA\right)
\right]
\ee
Matching coefficients of $\vphi$ gives
\be
\vA^\prime= \vA\vQ^\prime \vA
\label{eq:Aevolve}
\ee
\be
\vb^\prime = \vA\vQ^\prime \vb
\label{eq:bevolve}
\ee
\be
\sN^\prime = 
\frachalf
\vb^t
\vQp
\vb
+\frachalf {\rm Tr}\left[\vQ^{-1}\vQp \right]
-\frachalf {\rm Tr}\left[ \vA\vQp \right]
\label{eq:Nevolve}
\ee
For Gaussian fields, these are the main result equations in the paper. 
Note that we are 
assuming that, if $\vN$ is sufficiently short range for this approach to
work at all, computing $\vN^{-1}$ to set the initial conditions for this
evolution will not be a limitation. 

To gain understanding, 
we can unpack what the equation $\vA^\prime = \vA \vQ^\prime \vA$ is doing.
It is equivalent to $\partial \vA^{-1}/\partial \ln \lambda = -\vQ^\prime$, 
i.e., $\vA^{-1} = \vQ_0 - \vQ +\vA^{-1}_0 = \vP+\vN - (\vP^{-1}+\vK)^{-1}$ 
(remember that $\vA^{-1}_0=\vN$, where we are setting $\vR=\vI$ in this 
strictly pedagogical paragraph). 
I.e., $\vA^{-1}$ evolves from
$\vN$ when $\vK\rightarrow 0$ to $\vC=\vP+\vN$ when $\vK\rightarrow \infty$,
i.e., as the calculation runs we will have in some sense $\vA=\vN^{-1}$ 
for large scales but $\vA=\vC^{-1}$ for small scales. 
The evolution equation for $\vb$ has a simple solution in terms of
$\vA$: $\vb=\vA \vo$, i.e., given that we understand $\vA$ evolution as 
interpolation from $\vN^{-1}$ to $\vC^{-1}$, $\vb$ evolution is interpolation
from $\vN^{-1}$ data weighting to $\vC^{-1}$ weighting. 
Similarly, $\sN(\lambda)=\frachalf \vo^t \vA \vo+
\frachalf {\rm Tr}\ln\left(2\pi \vA^{-1}\right)+
\frachalf {\rm Tr}\ln\left(2\pi \vQ\right)$. 
Noting that $\vQ$ evolves from $\vP$ to 
$\vK^{-1}$, we see that this evolves from the initial normalization piece  
$\sN(\vK=0)=\frachalf \vo^t \vN^{-1} \vo+\frachalf {\rm Tr}\ln(2\pi \vN)+
\frachalf {\rm Tr}\ln(2\pi \vP)$ to the final answer
$\sN(\vK\rightarrow \infty)=
\frachalf \vo^t \vC^{-1} \vo+\frachalf {\rm Tr}\ln(2 \pi \vC)+
\frachalf{\rm Tr}\ln(2 \pi \vK^{-1})$ (where the $\vK$ part cancels against the
integration term in Eq. \ref{eq:infKintegral}). 
The whole point of the paper is that
you get to this final answer without actually evaluating these matrix 
formulas -- $\sN$ is just an
evolving single number, $\vb$ is an evolving vector. The key is keeping 
$\vA$ from spreading too far off-diagonal, 
because formally you do need to store is as a matrix. This is where the
coarsening step we will discuss below becomes necessary -- $\vA$ obviously
does  
spread off-diagonal, but as it does it becomes unnecessary to store it at high 
resolution, so we can coarse grain to keep the numerically relevant number of 
cells off-diagonal
limited, even as the physical range increases. 

Note that for the Gaussian problem we can always formally integrate Eq. 
\ref{eq:IderiveRG} over $\vphi$ to obtain
\be
L(\vtheta|\vo)=e^{\frachalf\vb^t \left(\vQ^{-1}+\vA\right)^{-1}\vb -\sN}
\sqrt{\det\left(2\pi \left(\vQ^{-1}+\vA\right)^{-1}\right)}~.
\label{eq:Qformintegral}
\ee
All of the components will change with $\lambda$, but the overall result will 
not change.
Once $\vK>>\vP^{-1}$, and $\vK>>\vA$, and
$\vb^t \vK^{-1}\vb\rightarrow 0$ (all of which we showed above inevitably do
happen as $\lambda\rightarrow \infty$) this goes to
\be
L(\vtheta|\vo)\xrightarrow{\lambda\rightarrow\infty}
e^{\frachalf\vb^t \vK^{-1}\vb -\sN}
\sqrt{\det\left(2\pi \vK^{-1}\right)}
\rightarrow e^{-\sN}
\sqrt{\det\left(2\pi \vK^{-1}\right)}~,
\label{eq:infKintegral}
\ee
i.e., the final
answer is just  $e^{-\sN}$, up to a $\det \vK$ factor which
is canceled by a similar factor
accumulated within $\sN$, as mentioned above. We have accomplished the
integration over $\vphi$. 

\subsection{Cutoff function }

Eqs. (\ref{eq:Aevolve}-\ref{eq:Nevolve}) 
are no good if we can't evaluate them numerically. 
The key is to arrange $\vQ^\prime$ to do its job of suppressing the smallest
scale structure to allow coarse graining, but beyond that have as short a real 
space range 
as possible (i.e., not coupling widely separated cells), 
to keep $\vA$, which 
starts as $\vR^t \vN^{-1}\vR$, as short range as possible 
(i.e., effectively sparse, when stored in real space).  
For the above definition we have
$\vQ=\left(\vP^{-1}\vW^{-1}+\vK\right)^{-1}$, so
\be
\vQp =\vQ \left(\vP^{-1}\vW^{-1} \vW^\prime\vW^{-1} -\vK^\prime\right)\vQ~.
\ee 
If $\vK=0$ this is just $\vQ^\prime = \vW^\prime\vP$. On the other hand, if
$\vW=1$ we have $\vQ^\prime =-\vQ\vK^\prime \vQ$.
Let's focus on the $\vW=1$ case, and assume $\vK$ will not depend on the
power spectrum as a function of parameters of interest (it might depend on 
a fiducial power spectrum that would be kept fixed when varying parameters). 
Making $\vK$ a simple function 
of $k$ in Fourier space, we can efficiently compute $\vQ^\prime$ in real 
space, which takes the form of a convolution kernel (meaning we don't need to
compute it as a full matrix, not even a sparse one -- we can compute it as
a function of separation and use translation invariance to move it around). 

We are left to choose a specific $\vK$. Suppose $\vP$ is a constant $P$ in 
Fourier space and we 
choose $K(k,\lambda)=P^{-1} k^2 \lambda^2 \exp(k^2 \lambda^2)$ which 
accomplishes the goal of dominating (going to $\infty$) at 
$k\lambda\gg 1$ and 
disappearing (going to zero)
at $k\lambda \ll 1$, then
$Q^\prime(x=k \lambda)=-2 P \frac{x^2 (1+x^2) 
\exp(x^2)}{\left(1+x^2 \exp(x^2)\right)^2}$. This is a fairly nice function
that goes like $x^2$ at small $x$, peaks at $x\sim 1$, and goes like 
$\exp(-x^2)$ at large $x$ --
it should produce a nice limited range convolution kernel.
Another very simple example is $K(k,\lambda)=P^{-1}(k) \lambda^2 k^2$. This 
gives $\vQ^\prime$ going
like $P(k) k^2$ at $k\lambda \ll 1$ and 
$P(k) k^{-2}$ at $k\lambda\gg 1$. We 
see that the function we choose is picking the detailed range of $P(k)$ to 
emphasize for a given $\lambda$ (when varying parameters, the $P(k)$ in 
$K$ would stay fixed).  
In the numerical example below we will use
$K(k,\lambda)=P_0^{-1}(k) (k \lambda)^\alpha \exp[(k \lambda)^2]$, 
with $\alpha=2$ or 2.5,
where $P_0(k)$ is the fiducial power spectrum. 
Tuning the cutoff function is a topic for more thought and experimentation.

\subsection{Coarse graining}

In any case, as $\lambda$ increases 
successive convolutions with $\vQ^\prime$ in general extend 
the correlation range of $\vA$, eventually making it the dense matrix 
($\vC^{-1}$ actually, as mentioned above)
that we want to avoid, 
if we don't do anything about it
(to be clear: evaluating 
Eq. \ref{eq:Aevolve} scales like the $N^3$ we said was too large if $\vA$
becomes dense). At the same time, however,
the growth of $\vK$ suppressing high-$k$ fluctuations in $\vphi$
makes resolution much finer than approximately the same scale unnecessary, 
i.e.,
we should be able to coarsen the pixelization to reduce the range of $\vA$  
as measured in cells off-diagonal. This can be seen by imagining Eq. 
(\ref{eq:Qformintegral})
in Fourier basis. At high $k$ where $\vK\rightarrow \infty$, any other 
structure is overwhelmed so the full simplification of 
Eq. (\ref{eq:infKintegral}) applies, 
i.e, apparently the values of $\vb$ and $\vA$ in these high-$k$ 
modes are irrelevant. 
All of the integral contribution in high-$k$ modes must now be contained in
$\sN$ -- this is where we see that the RG evolution equation is equivalent to 
integrating over these modes.
Since we know that high $k$ modes in $\vb$ and $\vA$ do no contribute
to the remaining integral, we can compress them to make the 
computations faster. 
Another way to see this is: 
all of the terms in the evolution equations 
(Eqs. \ref{eq:Aevolve}-\ref{eq:Nevolve}) involve
multiplication of $\vA$ or $\vb$ by $\vQ^\prime$, but by construction 
$\vQ^\prime$ has no structure on scales much less than $\lambda$
(i.e., is zero in Fourier space). This means
that, operationally, it is clear that $\vA$ and $\vb$ can be compressed
at some point, i.e.,
their small scale structure is irrelevant when it is multiplied by a 
smooth function. 
Imagining 1D, the compression process can be simply 
$b^c_1=b_1+b_2$, $b^c_2=b_3+b_4$, $A^c_{11}=A_{11}+2 A_{12}+A_{22}$, etc., 
where $\vb^c$ and $\vA^c$ are the new objects compressed by a factor of 2 in 
each index. Understanding that on small scales 
$\vb \rightarrow \vC^{-1}\vo$ and $\vA \rightarrow \vC^{-1}$, we see that this
process is intuitively straightforward, equivalent to an optimally weighted 
averaging of adjacent cells. 

Especially in higher dimensions, the problem is greatly simplified after
even a single 
factor of 2 in coarsening. Eventually the data set contains few 
enough cells to just evaluate the remaining integration by brute force
(i.e., plug $\vA$, etc. that we have at that point into 
Eq. \ref{eq:Qformintegral} and compute using standard linear algebra).
To parallelize the algorithm, we could split the data set by processor, where
for non-shared memory, each processor would need enough extra data around its
local patch to account for the convolution from outside.  

For a summary of the properties of various quantities we have defined, see 
Table \ref{tab:behavior}. 
\begin{table}
\begin{threeparttable}
\caption{Evolution of various quantities from start to asymptotic endpoint 
(for $\vR=\vI$). $\vP$ and $\vK$, and thus $\vQ$ and $\vQ^\prime$, are 
diagonal in Fourier space, while $\vN$ is at least approximately diagonal
in real space, and $\vA$ remains so after coarse graining (it would become
dense without coarse graining because $\vC\equiv \vP+\vN$ is dense). 
\label{tab:behavior}}
\begin{tabular}{lccc}
\hline
quantity & start ($k \lambda \ll 1$)& $\rightarrow$ & 
end ($k \lambda \gg 1$)  \\
\hline
$\lambda$ & 0 &  & $\infty$ \\
$\vK$ & 0 & & $\infty$ \\
$\vQ$ & $\vP$ & & $\vK^{-1}\rightarrow 0$ \\
$\vQ^\prime$ & 0 & & 0 \tnote{a}   \\
$\vA$ & $\vN^{-1}$ &  & $\vC^{-1}$ \tnote{b} \\
$\vb$ & $\vN^{-1}\vo$ &  & $\vC^{-1}\vo$ \tnote{b} \\
$\sN$ &  
$\frachalf \vo^t \vN^{-1} \vo+\frachalf {\rm Tr}\ln(2\pi \vN)+
\frachalf {\rm Tr}\ln(2\pi \vP)$ &  &
$\frachalf \vo^t \vC^{-1} \vo+\frachalf {\rm Tr}\ln(2 \pi \vC)$
\tnote{c}  \\
\hline
\end{tabular}
\begin{tablenotes}
\item [a] non-zero for $k\lambda\sim 1$, 
see Figs. \ref{fig:dQk} and \ref{fig:dQr}
\item [b] coarse grained
\item [c] $\det \vK$ term canceled 
\end{tablenotes}
\end{threeparttable}
\end{table}

\subsection{Numerical example}

\begin{figure}[ht!]
 \begin{center}
  \includegraphics[scale=1.0]{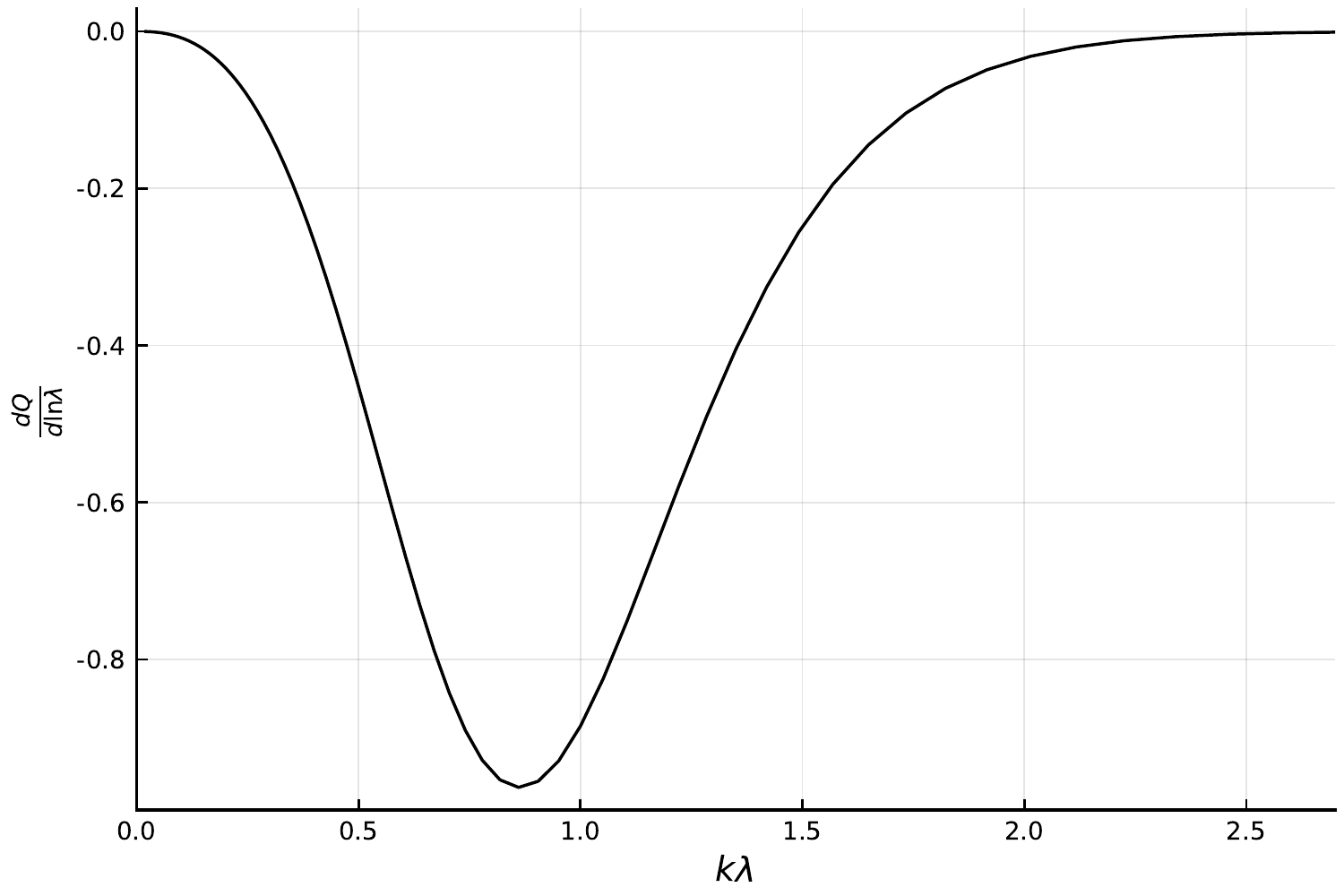}
 \end{center}
 \caption{ 
Fourier space $\frac{dQ}{d\ln \lambda}(k)$ vs. $k\lambda$. 
This shows the range in which
fluctuations in $\vphi$ are being ``integrated out''.  
 }
 \label{fig:dQk}
\end{figure}

\begin{figure}[ht!]
 \begin{center}
  \includegraphics[scale=1.0]{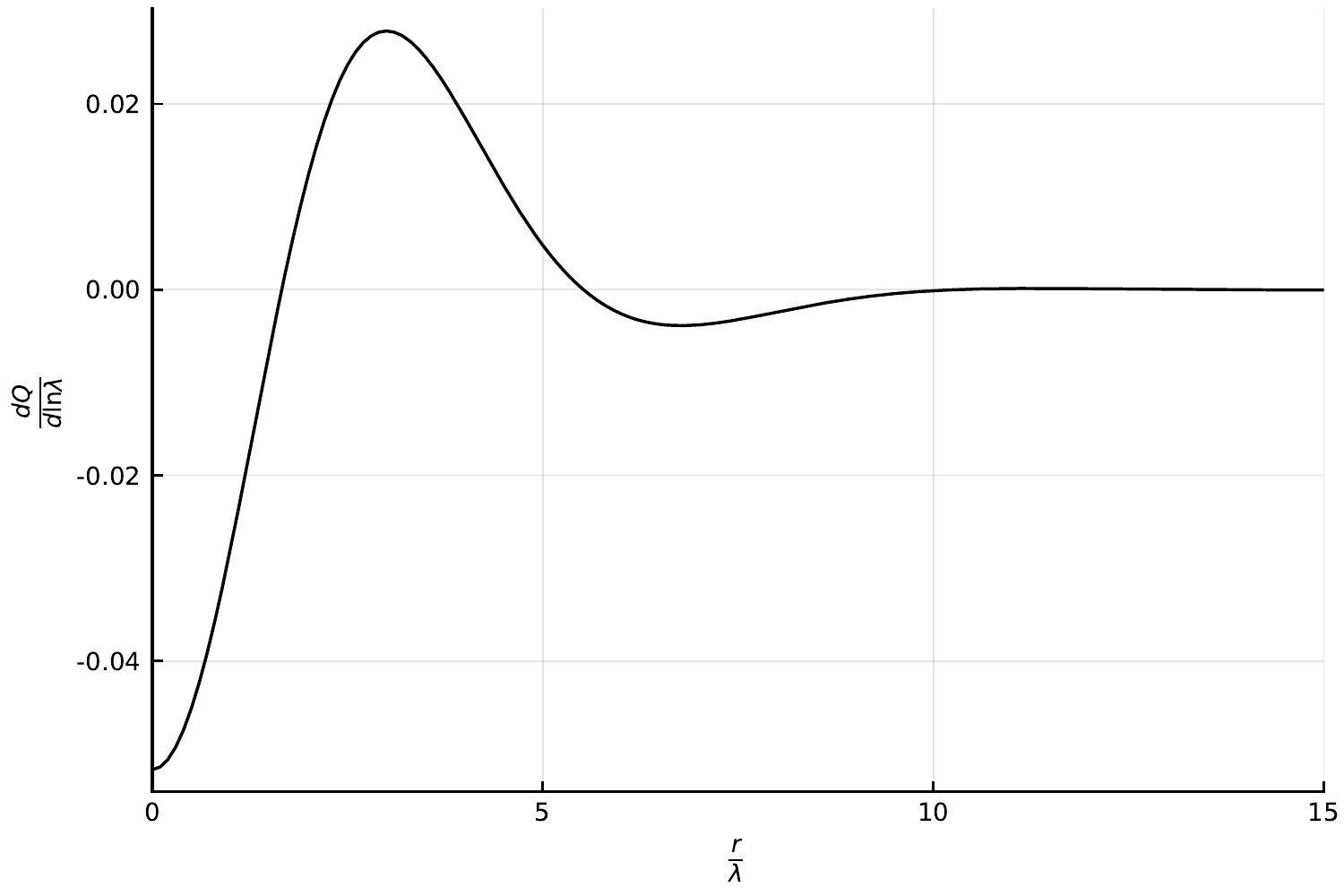}
 \end{center}
 \caption{ 
Real space $\frac{dQ}{d\ln \lambda}(r)$ vs. $\frac{r}{\lambda}$. 
This shows the convolution kernel applied during the evolution of 
$\vA$, $\vb$, $\sN$.
}
 \label{fig:dQr}
\end{figure}

Eqs. (\ref{eq:Aevolve}-\ref{eq:Nevolve}) are exact for the Gaussian problem, so
in some sense they must work, but it is helpful to work through a numerical 
example to reassure ourselves that they are practical. For coding 
simplicity, I work in one dimension, in units of the basic data cell size.  
I include unit variance noise in each cell. 
I use signal power spectrum $P(k)=A (k/k_p)^\gamma \exp(-k^2)$ 
with $\gamma=0$ or $-0.5$. 
I generate mock data with $A_0=1$ and 
calculate likelihoods as a function of $A$.
I use pivot $k_p=0.1$ to make the
constraining power on $A$ roughly similar between the two $\gamma$s (for my
largest datasets). 
I choose the cutoff
function $K(k,\lambda)=P_0^{-1}(k) (k \lambda)^\alpha \exp[(k \lambda)^2]$, 
where $P_0(k)$ is the fiducial power spectrum.
I initially used $\alpha=2$, but found that $\alpha=2.5$ seems to be
somewhat more efficient for the $\gamma=-0.5$ case. 
In Figure \ref{fig:dQk} I show the critical $\vQ^\prime$ in Fourier space (for
$P(k)=1$, to isolate the scale dependence of $\vK$ without mixing it with
$P(k)$). 
For any $\lambda$, this shows the range of $k$ that 
is being ``integrated out.'' The real space version of $\vQ^\prime$ is shown in 
Figure \ref{fig:dQr}.
It has mean zero, 
meaning it does not affect scales much larger
than $\lambda$ -- this is what keeps $\vA$ short range -- and it is smooth so 
that it does not affect scales much smaller than $\lambda$ -- this is why 
we can coarsen $\vA$ and $\vb$. 

\begin{figure}[ht!]
 \begin{center}
  \includegraphics[scale=1.0]{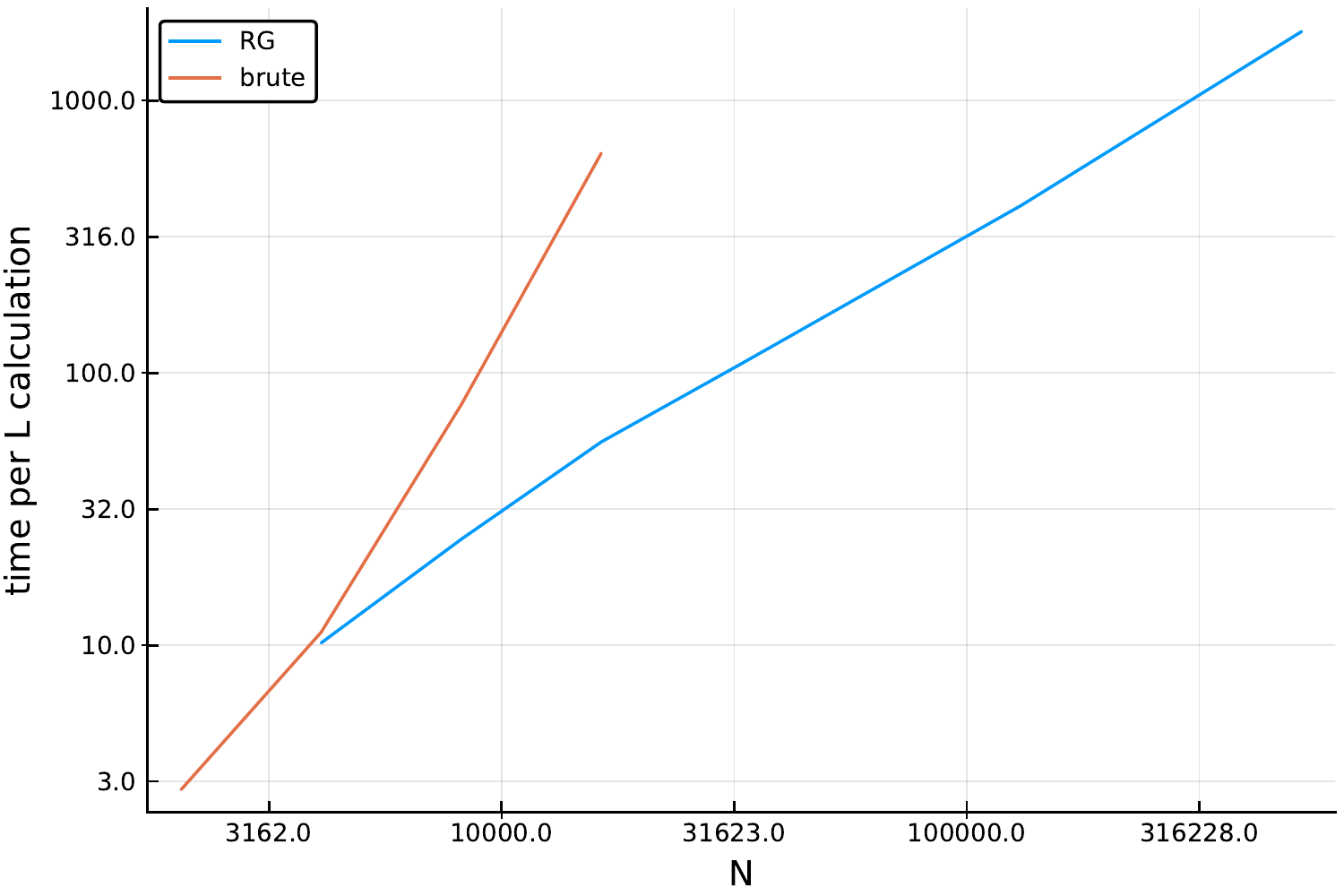}
 \end{center}
 \caption{
Time per likelihood evaluation, scaling linearly with $N_o$ for the RG method, 
compared to $N_o^3$ for brute force linear algebra. 
This is implemented using 
canned Julia sparse matrix routines, without much effort to optimize, so the
normalization can probably be reduced significantly. This is running on 1 
processor, where, e.g., the $N_o=524288$ run takes 30 minutes. 
}
 \label{fig:timing}
\end{figure}

For small problems, we can evolve Eqs. (\ref{eq:Aevolve}-\ref{eq:Nevolve}) 
using a dense matrix for $\vA$, and check step by step that the 
components and final likelihood agree with the analytic solutions.
I find that steps $d \ln \lambda=0.2$ are sufficient in practice for my
test problems, after realizing that it is very helpful to use at least 
2nd order 
accurate mid-point stepping 
(\url{https://en.wikipedia.org/wiki/Midpoint_method}), 
instead of the most naive possible 1st order 
stepping. For larger problems I use canned Julia sparse matrix operations for
$\vA$. Since $\vQ^\prime$ is formally dense (just falling rapidly to zero with 
separation), we need to impose by hand a cut on the range of matrix operations 
to store, which I set after experimentation to $20 \lambda$. Note that this,
like $d \ln \lambda$ and other numerical parameters of the method, is a highly
controlled approximation that can be checked for convergence without reference
to any known answer. The final non-trivial numerical parameter is when to 
coarsen, which I set after experimentation to when $\lambda$ has evolved to
be 7 times the current cell size, i.e., the formula is 
``if $\lambda>7 c$, $c \rightarrow 2 c$'', where $c$ is the cell size. 
These settings determine the overall speed 
of the analysis, because they determine how many cells off diagonal must be
stored in $\vA$, i.e., the product of the two numbers tells us how efficiently
we are accomplishing the dual goals of rubbing out small-scale structure enough
to coarsen without broadening the range at which we need to store 
correlations any more than necessary. 
It may seem like my numbers are surprisingly large, but I
am requiring quite high accuracy in the likelihood calculation, and 
did not try very hard to, e.g., tune the cutoff function to potentially 
reduce them. 
Fig. \ref{fig:timing} shows the time per likelihood evaluation, scaling like
$N_o$, compared to brute force linear algebra scaling like $N_o^3$. 
Once the data set is coarsened to 2048 cells I simply evaluate the remaining
integral all at once using Eq. (\ref{eq:Qformintegral}). 

\begin{figure}[ht!]
 \begin{center}
  \includegraphics[scale=1.0]{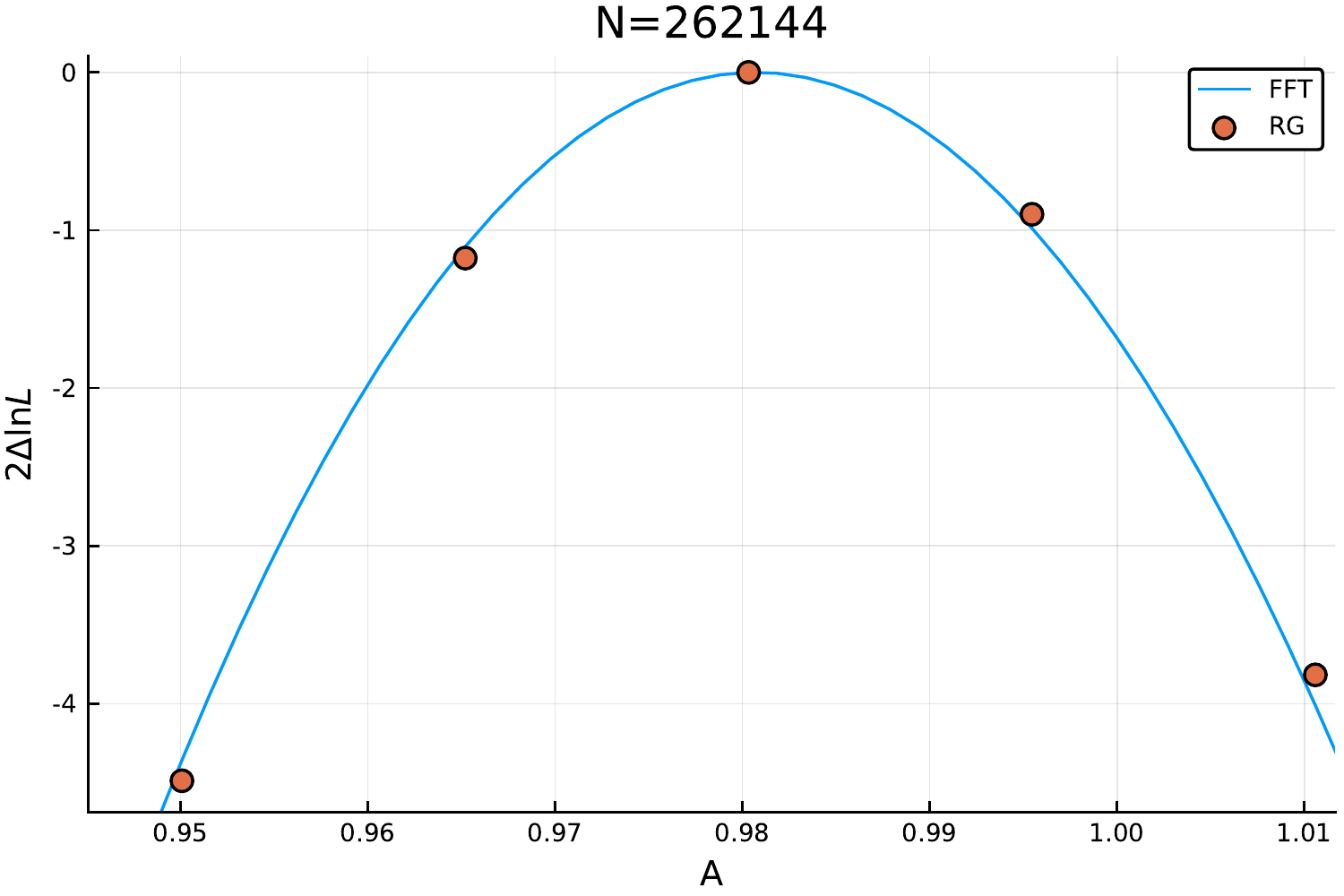}
 \end{center}
 \caption{
RG likelihood for five values of the power amplitude, compared to the exact
FFT likelihood that I can compute because this data set was periodic with 
homogeneous noise. This is for $\gamma=-0.5$, $N=262144$ data points. 
All similar figures look
similar, by construction, because if they didn't I would have dialed up the 
numerical accuracy parameters until they did. 
}
 \label{fig:lnL}
\end{figure}

At this point plotting the results is somewhat anti-climactic, because the 
calculation simply works, more or less by construction. 
Fig. \ref{fig:lnL} shows the likelihood computed for an $N_o=262144$ 
data set for different power amplitude
values, compared to an exact FFT likelihood that I can compute for this 
data set because I made it periodic with homogeneous noise. 
The RG calculation does not take advantage of this symmetry 
in any way. There is one
imperfection hidden in Fig. \ref{fig:lnL}, which shows not absolute 
$2 \ln L$ but its deviation from maximum -- the maximum is off by 36 in the RG 
calculation. This is like if you were doing a $\chi^2$ test of goodness of
fit, and got $\chi^2$ wrong by 36 for 262144 degrees of freedom -- it is 
insignificant to the goodness of fit calculation, which has an rms 
error of 724. The important thing is that the offset is practically the same
for all parameter values, so does not affect our conclusions about the 
parameter (I can make $\chi^2$ more accurate by dialing up numerical settings, 
but there is no 
point \footnote{One may wonder how generic is this behavior that 
differences in $\chi^2$ between models have much better accuracy 
than the 
absolute value of $\chi^2$. In practice it does not really matter because we
will always test numerical convergence of the quantities of interest
(generally differences between models) on a case-by-case basis, determining
whether they converge without regard to whether they converge faster than 
absolute $\chi^2$ or not, but maybe it is an interesting academic question.
I think this behavior probably is generic, because of the usual relation 
between absolute $\chi^2$, degrees of freedom (dof), 
and $\Delta \chi^2$ between 
models. Absolute $\chi^2$ for any data set has big random fluctuations  
around a 
big number dof, much more for a large data set than the level of 
$\Delta \chi^2$ we would call significant as a difference between models 
(e.g., we would say a model is ruled out decisively if it fits 
$\Delta \chi^2=25$ worse than another one, even if both of them would look 
like they had a ``fine $\chi^2$'', i.e., $\chi^2\sim$ dof, if compared to that 
data in isolation). The 
reason these things work this way is that most of the contribution to absolute
$\chi^2$ is effectively orthogonal to any smooth model, so it is irrelevant 
to comparison between models. I think we can generically expect this 
likelihood calculation to behave similarly. 
Again though, we can/must always do convergence tests to verify that the 
likelihood differences that we care about are not sensitive to numerical 
parameters, so we won't ever be fooled by imperfection in this intuition.}).

While it is pretty clear that the method will deal gracefully with edges and
varying noise, just to be sure I did some tests of this with $N_o=4096$ data 
sets, for which I can calculate the exact answer by brute force. One test 
was to multiply the noise in cells in the second half of the data vector by a 
factor of 10, and the other was to multiply the noise in even-index 
(i.e., every other) cells by 10. These tests give similar agreement to 
what we see in Fig. \ref{fig:lnL}. 
Note that unobserved cells are equivalent to infinite noise, which enter these
calculations nicely as $N_{ii}^{-1}=0$, which can be dropped from sparse 
storage of $\vA$. Formally infinite ``zero padding'' around an
observed region, i.e.,  
consistently implemented unobserved cells, incur no computational cost
beyond the range of smearing of $\vA$. 

There are some subtleties in numerical evaluation that a potential user should
be aware of. You don't want to compute the quantity 
$\left(\vQ^{-1}+\vA\right)^{-1}$ that appears in 
Eq. \ref{eq:Qformintegral} naively
by computing $\vQ^{-1}$, adding that to $\vA$, and inverting the resulting 
matrix, because
$\vQ^{-1}\rightarrow \vK\rightarrow \infty$ for high $k$ modes, meaning in 
practice that, after some evolution, there are elements of $\vQ^{-1}$ too big 
for the computer to handle. The formally equivalent quantity 
$\vQ\left(\vI+\vA\vQ\right)^{-1}$ is well behaved when used to compute
$\vb^t \left(\vQ^{-1}+\vA\right)^{-1}\vb$, because the
high $k$ elements of $\vQ$ go to zero.
This trick does not entirely solve the problem for 
$\det\left[\left(\vQ^{-1}+\vA\right)^{-1}\right]$, however because 
near-zero eigenvalues are a problem for the determinant.  
At this point we notice, however, that we don't need or want to actually
include the term  ${\rm Tr}\left[\vQ^{-1}\vQp \right]$ in our 
numerical 
evolution of $\sN^\prime$ (Eq. \ref{eq:Nevolve}), 
because we can use the analytic 
solution of this piece of the equation, 
${\rm Tr}\ln \vQ$, to analytically cancel the problematic 
$\det{\vQ}$ in 
$\det\left[\vQ\left(\vI+\vA\vQ\right)^{-1}\right]$ 
in Eq. \ref{eq:Qformintegral}, leaving us to compute the 
perfectly well-behaved $\det\left[\vI+\vA\vQ\right]$. 

\section{Beyond-Gaussian likelihood  \label{sec:nonG}}

The derivation of Eq. (\ref{eq:masterRG}) made no assumption about the form
of $\phi$ dependence of $S_I$ -- it is valid for general $S_I(\phi)$. 
In practice $S_I$ can be something other than quadratic due to primordial 
non-Gaussianity \cite[e.g.,][]{2008PhRvD..78l3519M,
2010JCAP...03..011B,
2010PhRvD..81f3530G,2011MNRAS.417L..79G,2014arXiv1412.4671A,
2018JCAP...01..010M}, non-linear gravitational evolution
\cite[e.g.,][]{
2001A&A...379....8V,
2002A&A...382..412V,
2004A&A...421...23V,
2007A&A...465..725V,
2007PhRvD..75d3514M,
2011JCAP...04..032M,
2011JCAP...11..039S,
2012JCAP...07..051B,
2012JHEP...09..082C,
2014PhRvD..89d3521H,
2014arXiv1403.7235M,
2015JCAP...11..049B,
2015JCAP...01..014R,
2016JCAP...02..032F,
2016NJPh...18d3020B,
2016PhRvD..93j3528S,
2016MNRAS.463.2273F,
2018PhRvD..97b3508M,
2018JCAP...09..006J,
2018JCAP...06..028P,
2018JCAP...05..039P,
2018JSMTE..04.3214F,
2017PhRvD..96l3532K,
2018arXiv180704215T}, gravitational lensing
\cite[e.g.,][]{2014JCAP...09..024J,2015JCAP...01..022M,2017JCAP...03..016D},  
non-linear bias \cite[e.g.,][]{
2006PhRvD..74j3512M,
2007PhRvD..75f3512S,
2008PhRvD..78h3519M,
2009JCAP...08..020M,
2009ApJ...691..569J,
2011PhRvD..83h3518M,
2011MNRAS.416.1703E,
2012PhRvD..86j3519C,
2012PhRvD..86h3540B,
2014MNRAS.444.1400N,
2014PhRvD..90l3522S,
2016arXiv161109787D,
2017arXiv170602362H,
2017PhRvD..95f3528C,
2017MNRAS.467.3993A}, 
non-Gaussian noise \cite[e.g.,][]{2009PhRvD..80j5005E}, etc. 
\cite[e.g.,][]{2017JCAP...03..006D}.
Eq. (\ref{eq:masterRG}) always applies formally,
but the tricky part is that we must be able to match the functional dependence 
generated on the right hand side  
against terms on the left hand side. 
E.g., if we 
add a $\phi^3$ term to $S_I(\phi)$, we will find that all higher order 
polynomial terms are generated by the evolution. This is due to the 
$\frac{\partial S_I} {\partial \vphi^t} \vQp
\frac{\partial S_I} {\partial \vphi}$ term in Eq. (\ref{eq:masterRG}), which
feeds polynomial terms of order $n$ and $m$ into a term of order 
$n+m-2$, i.e., if $n, m\leq 2$, this is never larger than 2, but
$n=m=3$ leads to a fourth order term, and so on.
(Fig. 9.2 of \cite{2008mqft.book.....B} 
nicely represents Eq. \ref{eq:masterRG} in terms of Feynman diagrams.) 
These higher order terms aren't necessarily 
relevant though, i.e., computationally significant, so one can hope to simply
ignore terms beyond some order. Intuitively, this should be more likely to work
if one starts the evolution at relatively low resolution, i.e., on larger, more 
perturbative scales. Note that a polynomial $S_I$ is not necessarily 
the best way to represent non-linearity \cite[e.g.,][]{2018PhRvD..97b3508M}.

\section{Derivatives of $\ln L$, i.e., quadratic estimator-type application}

We could always run the above likelihood calculation 
for different parameter values in order
to compute parameter dependence, however, it may be more efficient to 
derive the analytic formula for derivatives with respect to parameters 
and then compute that from the
data, in the usual way of ``quadratic estimators.'' 
Derivatives with respect to parameters bring down quadratic factors in
$\vphi$. 
We should be able to similarly accumulate these integrals. 

Consider
\be
\frac{\partial \ln L(\vtheta|\vo)}{\partial \theta_\alpha}= 
-\frachalf{\rm Tr}\left[\vP^{-1}\vP_{,\alpha} \right]+
\frachalf\frac{ 
\int d\vphi~ 
\vphi^t \vP^{-1}\vP_{,\alpha} \vP^{-1} \vphi
~e^{-\frachalf \vphi^t \vP^{-1} \vphi
- S_I(\vphi)}}{
\int d\vphi~ e^{-\frachalf \vphi^t \vP^{-1} \vphi
- S_I(\vphi)}}{
}
\label{eq:firstderiv}
\ee
where $\vP_{,\alpha}\equiv \partial \vP/\partial \theta_\alpha$. Note that
we do not need to keep track of the common normalization factor between 
the numerator and denominator. We now look to accumulate the 
contribution coming from insertion of 
$\frachalf \vphi^t \vP^{-1}\vP_{,\alpha} \vP^{-1} \vphi$. The leading 
trace factor can be computed trivially in Fourier space.

Noting that
$f[\vphi]~e^{-S(\vphi)} =e^{ - S(\vphi)+\ln f[\vphi]}$
we see that we can compute integrals 
like in Eq. (\ref{eq:firstderiv}), i.e., 
integrals of the form 
$\int d\vphi~ f[\vphi] ~e^{-\frachalf \vphi^t \vP^{-1} \vphi - S_I(\vphi)}$,
by simply inserting  $S_I-\ln f$ in place of 
$S_I$ in Eq. (\ref{eq:masterRG}).
In this way we derive, in addition to the same evolution equations for $S_I$ 
(i.e., $\vA$ and $\vb$) above,
\be
f^\prime =
\frac{\partial f}{\partial \vphi^t}\vQ^\prime 
\frac{\partial S_I}{\partial \vphi}
-\frachalf {\rm Tr}\left[
\frac{\partial^2 f}{\partial \vphi \partial \vphi^t} \vQ^\prime\right]~.
\ee
Note that this is effectively a special case of the general non-Gaussianity
discussed in \S\ref{sec:nonG}, where the particular form of effective 
non-Gaussianity here guarantees that we have an exact basis to project onto.

In the case of Eq. (\ref{eq:firstderiv}), we have $f_\alpha[\vphi,\lambda]\equiv
\frachalf \vphi^t \vP^{-1}\vP_{,\alpha} \vP^{-1}\vphi+
\frachalf\vphi^t \vD_\alpha(\lambda)\vphi+
\vc_\alpha(\lambda)\vphi+
E_\alpha(\lambda)$,
where 
$\vD_\alpha(\lambda=0)=0$, $\vc_\alpha(\lambda=0)=0$, and 
$E_\alpha(\lambda=0)=0$, giving
\begin{eqnarray}
\frachalf\vphi^t \vD_\alpha^\prime\vphi+
\vc_\alpha^\prime\vphi+
E_\alpha^\prime 
&=&
\left(\vphi^t \vP^{-1}\vP_{,\alpha}\vP^{-1}+ \vphi^t \vD_\alpha +\vc_\alpha^t
\right) 
\vQ^\prime 
\left(\vA \vphi -\vb\right)
-\frachalf {\rm Tr}\left[ \vP^{-1}\vP_{,\alpha}\vP^{-1}\vQ^\prime\right]
-\frachalf {\rm Tr}\left[ \vD_\alpha\vQ^\prime\right]
\end{eqnarray}
Matching coefficients of powers of $\vphi$ we find:
\be
\vD^\prime_\alpha = 
\vP^{-1}\vP_{,\alpha}\vP^{-1} \vQ^\prime \vA+
\vD_{\alpha}\vQ^\prime \vA+{\rm transpose}
\ee
(``transpose'' here means transpose of the part I wrote on the right hand side
to make $\vD_\alpha^\prime$ symmetric)
\be
\vc_\alpha^\prime=\vA\vQ^\prime \vc_\alpha -
\vP^{-1}\vP_{,\alpha}\vP^{-1} \vQ^\prime \vb
-\vD_{\alpha}\vQ^\prime \vb
\ee 
and finally
\be
E^\prime_\alpha = -\vc_\alpha^t \vQ^\prime \vb 
-\frachalf
{\rm Tr}\left[\vP^{-1}\vP_{,\alpha}\vP^{-1} \vQ^\prime\right]
-\frachalf
{\rm Tr}\left[\vD_{\alpha} \vQ^\prime\right]
\ee
Note that once $\lambda\rightarrow \infty$, all expectation values of 
$\vphi$ go to zero, so the final result of interest is just
$E_\alpha$ (which comes with a normalization factor that is canceled
by the denominator). 
The accumulation of $\vD_\alpha$, $\vc_\alpha$, and $E_\alpha$ will 
be dominated by the range of $\lambda$ corresponding to their band scale, so
it should not be necessary to compute them for all $\lambda$, i.e., for a given
$\lambda$ only a limited range of bands near this scale will need to be 
computed, so the computation time should not increase by too large a factor
over a single likelihood calculation.

The key to the efficiency of this calculation appears to be
keeping $\vD_\alpha$ short range, as we discussed for $\vA$ above. 
The key to this is the quantity $\vP^{-1}\vP_{,\alpha}\vP^{-1} \vQ^\prime$ 
being a short-range convolution. The $\vQ^\prime$ is no more problem than
discussed above, but $\vP_{,\alpha}$ brings up a new worry. If this represents
a band in power narrower than $\vQ^\prime$, 
the range of convolution must be inversely
proportional to the width of the band, i.e., 
increasing $k$ resolution in the power spectrum measurement will generally
require extended range of $\vD_\alpha$. As long as the required
fractional resolution is reasonable, this will just be a
prefactor on the fast $\mathcal{O}(N)$ scaling.  

The 2nd derivative of $\ln L$ will proceed similarly, with two-index
objects to accumulate. 

\section{Discussion}

Obvious next steps include applying this approach to real data analysis, and
thinking further beyond Gaussianity. 
Note that the definition of $\vQ$ does not need to be isotropic. E.g., for a
spectroscopic redshift survey like DESI overlapping with a lensing survey
like LSST \cite{2014JCAP...05..023F,2016arXiv161100036D},
we may want to first integrate out small-scale radial modes to 
which the lensing survey is not sensitive, before transitioning to a joint
analysis of primarily transverse modes. $\vQ$ can and should be tuned to the
problem at hand, with the only requirement being that calculations are 
tractable and have the right limits (i.e, $\vQ$ must start as $\vP$ and go to 
zero for any mode that you want to compress away). Note 
that in greater than one dimension 
it may be faster to integrate and compress by factors of 2 in one 
dimension at a time instead of all simultaneously. It might also be possible
to compress by less than factors of 2, interpolating onto an intermediate 
resolution grid instead of simply adding perfectly aligned pairs of cells. 

Considering Fig. \ref{fig:dQr}, it seems likely that a more carefully chosen
function that had, in particular, only one zero-crossing, could be more 
efficient. I.e.,
we need one zero-crossing to have mean zero, but given that we would like the
function to be as smooth and compact as possible, and two zero-crossings
surely aren't optimal. 
Now that we understand the method better, we see that 
it may be simpler to use the multiplicative $\vW$ form of cutoff instead
of the additive $\vK$. 
Another possibility might be
to give up exact translation invariance in $\vW$ or $\vK$ and instead choose 
them to specifically suppress the difference between cells one wants to sum
(as opposed to all nearby cells, which is what the translation invariant 
version does, but may be inefficient). 

For the Gaussian problem where we can solve the evolution equations 
analytically (see \S\ref{sec:GaussianRG}), there is an interesting possibility
of jumping directly from one coarsening step to the next using 
$\vA^{-1}=\vA_0^{-1} +\vQ_0-\vQ$, i.e., given $\vA_0^{-1}$ after a coarsening, 
we take the full $\lambda$ step to the next coarsening by computing
$\vQ_0-\vQ$ and simply adding -- this quantity will have similar sparsity
properties to what we have discussed for $\vA$. The trade-off appears to be
that now to propagate the contribution to $\sN$ we will need to numerically 
invert and compute the determinant of this $\vA^{-1}$. 
This should be doable because it is sparse, but may not be any faster than 
just taking the differential steps as in this paper. Note that I only took
3-4 steps between factors of 2 coarsening in the example problem, 
with the exception being leading up to the first coarsening (and even there,
it is the last steps before the coarsening that take the most time, because
$\vA$ has spread the most). This idea would probably be most interesting if 
in some scenario it is found that significantly finer $\ln \lambda$ steps are 
required. 

This type of RG approach has been applied to calculations of 
data-independent statistics of the evolving density field
\cite[e.g.,][]{2007JCAP...06..026M,2007PhRvD..76h3517I,
2008JCAP...01..029R,2008MPLA...23...25M,2017JCAP...01..048F}, 
but the very simple 
application here may cast new light on how to think about those 
applications. \cite{2018PhRvD..97b3508M} provides an exact functional 
integral starting point for this.  
This RG approach should also be another tool within the
``information field theory'' formalism of  
\cite{2009PhRvD..80j5005E,2018arXiv180403350E}.

A fully dense, arbitrary, inhomogeneous noise matrix will of 
course confound this 
method. A translation invariant noise component, i.e., diagonal in Fourier 
space, could be included 
in the signal power spectrum part of the calculation. A contaminant that can
be modeled by a reasonable number of vectors contributing to the mean of
the field can be marginalized over (with simplifications if the vectors are,
e.g., smooth or localized).

Finally, 
asking the question ``what will we do with a redshift survey of the real, 
evolving (i.e., not translation invariant) Universe'' leads to the following
re-understanding of the key points of the method:
Take the definition $\vQ\equiv \vW\vS$, where I use $\vS$ here where I had 
used $\vP$ before
to emphasize that the signal covariance matrix does not need to be
neatly diagonalizable by a Fourier transform. $\vW$ starts at $\vI$, while 
we can remove a mode from the data set if $\vW$ drives its variance to zero. 
This is simplest to visualize and implement if $\vS$ and $\vW$ can be 
multiplied along the
diagonal in Fourier space, but clearly a $\vW$ matrix representing any kind of 
smoothing operation that suppresses small-scale fluctuations 
will allow us to coarse grain. The requirement for the
calculation to work is simply that we can efficiently compute a compact
$\vQp=\vW^\prime \vS$.  

\acknowledgments

I thank U. Seljak, S. Ferraro, E. Castorina, and especially Alex Kim
for helpful comments on the draft. 

\bibliography{cosmo,cosmo_preprints}

\begin{thebibliography}{72}%
\makeatletter
\providecommand \@ifxundefined [1]{%
 \@ifx{#1\undefined}
}%
\providecommand \@ifnum [1]{%
 \ifnum #1\expandafter \@firstoftwo
 \else \expandafter \@secondoftwo
 \fi
}%
\providecommand \@ifx [1]{%
 \ifx #1\expandafter \@firstoftwo
 \else \expandafter \@secondoftwo
 \fi
}%
\providecommand \natexlab [1]{#1}%
\providecommand \enquote  [1]{``#1''}%
\providecommand \bibnamefont  [1]{#1}%
\providecommand \bibfnamefont [1]{#1}%
\providecommand \citenamefont [1]{#1}%
\providecommand \href@noop [0]{\@secondoftwo}%
\providecommand \href [0]{\begingroup \@sanitize@url \@href}%
\providecommand \@href[1]{\@@startlink{#1}\@@href}%
\providecommand \@@href[1]{\endgroup#1\@@endlink}%
\providecommand \@sanitize@url [0]{\catcode `\\12\catcode `\$12\catcode
  `\&12\catcode `\#12\catcode `\^12\catcode `\_12\catcode `\%12\relax}%
\providecommand \@@startlink[1]{}%
\providecommand \@@endlink[0]{}%
\providecommand \url  [0]{\begingroup\@sanitize@url \@url }%
\providecommand \@url [1]{\endgroup\@href {#1}{\urlprefix }}%
\providecommand \urlprefix  [0]{URL }%
\providecommand \Eprint [0]{\href }%
\providecommand \doibase [0]{http://dx.doi.org/}%
\providecommand \selectlanguage [0]{\@gobble}%
\providecommand \bibinfo  [0]{\@secondoftwo}%
\providecommand \bibfield  [0]{\@secondoftwo}%
\providecommand \translation [1]{[#1]}%
\providecommand \BibitemOpen [0]{}%
\providecommand \bibitemStop [0]{}%
\providecommand \bibitemNoStop [0]{.\EOS\space}%
\providecommand \EOS [0]{\spacefactor3000\relax}%
\providecommand \BibitemShut  [1]{\csname bibitem#1\endcsname}%
\let\auto@bib@innerbib\@empty
\bibitem [{\citenamefont {{Peebles}}(1993)}]{1993ppc..book.....P}%
  \BibitemOpen
  \bibfield  {author} {\bibinfo {author} {\bibfnamefont {P.~J.~E.}\
  \bibnamefont {{Peebles}}},\ }\href@noop {} {\emph {\bibinfo {title}
  {Principles of Physical Cosmology by P.J.E.~Peebles.~Princeton University
  Press, 1993.~ISBN: 978-0-691-01933-8}}}\ (\bibinfo {year} {1993})\BibitemShut
  {NoStop}%
\bibitem [{\citenamefont {{Bond}}\ \emph {et~al.}(1998)\citenamefont {{Bond}},
  \citenamefont {{Jaffe}},\ and\ \citenamefont {{Knox}}}]{1998PhRvD..57.2117B}%
  \BibitemOpen
  \bibfield  {author} {\bibinfo {author} {\bibfnamefont {J.~R.}\ \bibnamefont
  {{Bond}}}, \bibinfo {author} {\bibfnamefont {A.~H.}\ \bibnamefont {{Jaffe}}},
  \ and\ \bibinfo {author} {\bibfnamefont {L.}~\bibnamefont {{Knox}}},\
  }\bibfield  {title} {\enquote {\bibinfo {title} {{Estimating the power
  spectrum of the cosmic microwave background}},}\ }\href@noop {} {\bibfield
  {journal} {\bibinfo  {journal} {\prd}\ }\textbf {\bibinfo {volume} {57}},\
  \bibinfo {pages} {2117--2137} (\bibinfo {year} {1998})}\BibitemShut {NoStop}%
\bibitem [{\citenamefont {{Pen}}(2003)}]{2003MNRAS.346..619P}%
  \BibitemOpen
  \bibfield  {author} {\bibinfo {author} {\bibfnamefont {Ue-Li}\ \bibnamefont
  {{Pen}}},\ }\bibfield  {title} {\enquote {\bibinfo {title} {{Fast power
  spectrum estimation}},}\ }\href {\doibase 10.1046/j.1365-2966.2003.07118.x}
  {\bibfield  {journal} {\bibinfo  {journal} {\mnras}\ }\textbf {\bibinfo
  {volume} {346}},\ \bibinfo {pages} {619--626} (\bibinfo {year} {2003})},\
  \Eprint {http://arxiv.org/abs/astro-ph/0304513} {arXiv:astro-ph/0304513
  [astro-ph]} \BibitemShut {NoStop}%
\bibitem [{\citenamefont {{Padmanabhan}}\ \emph {et~al.}(2003)\citenamefont
  {{Padmanabhan}}, \citenamefont {{Seljak}},\ and\ \citenamefont
  {{Pen}}}]{2003NewA....8..581P}%
  \BibitemOpen
  \bibfield  {author} {\bibinfo {author} {\bibfnamefont {N.}~\bibnamefont
  {{Padmanabhan}}}, \bibinfo {author} {\bibfnamefont {U.}~\bibnamefont
  {{Seljak}}}, \ and\ \bibinfo {author} {\bibfnamefont {U.~L.}\ \bibnamefont
  {{Pen}}},\ }\bibfield  {title} {\enquote {\bibinfo {title} {{Mining weak
  lensing surveys.}}}\ }\href@noop {} {\bibfield  {journal} {\bibinfo
  {journal} {New Astronomy}\ }\textbf {\bibinfo {volume} {8}},\ \bibinfo
  {pages} {581--603} (\bibinfo {year} {2003})}\BibitemShut {NoStop}%
\bibitem [{\citenamefont {{Wandelt}}\ \emph {et~al.}(2004)\citenamefont
  {{Wandelt}}, \citenamefont {{Larson}},\ and\ \citenamefont
  {{Lakshminarayanan}}}]{2004PhRvD..70h3511W}%
  \BibitemOpen
  \bibfield  {author} {\bibinfo {author} {\bibfnamefont {Benjamin~D.}\
  \bibnamefont {{Wandelt}}}, \bibinfo {author} {\bibfnamefont {David~L.}\
  \bibnamefont {{Larson}}}, \ and\ \bibinfo {author} {\bibfnamefont {Arun}\
  \bibnamefont {{Lakshminarayanan}}},\ }\bibfield  {title} {\enquote {\bibinfo
  {title} {{Global, exact cosmic microwave background data analysis using Gibbs
  sampling}},}\ }\href {\doibase 10.1103/PhysRevD.70.083511} {\bibfield
  {journal} {\bibinfo  {journal} {\prd}\ }\textbf {\bibinfo {volume} {70}},\
  \bibinfo {eid} {083511} (\bibinfo {year} {2004})},\ \Eprint
  {http://arxiv.org/abs/astro-ph/0310080} {arXiv:astro-ph/0310080 [astro-ph]}
  \BibitemShut {NoStop}%
\bibitem [{\citenamefont {{Smith}}\ \emph
  {et~al.}(2007{\natexlab{a}})\citenamefont {{Smith}}, \citenamefont {{Zahn}},\
  and\ \citenamefont {{Dor{\'e}}}}]{2007PhRvD..76d3510S}%
  \BibitemOpen
  \bibfield  {author} {\bibinfo {author} {\bibfnamefont {Kendrick~M.}\
  \bibnamefont {{Smith}}}, \bibinfo {author} {\bibfnamefont {Oliver}\
  \bibnamefont {{Zahn}}}, \ and\ \bibinfo {author} {\bibfnamefont {Olivier}\
  \bibnamefont {{Dor{\'e}}}},\ }\bibfield  {title} {\enquote {\bibinfo {title}
  {{Detection of gravitational lensing in the cosmic microwave background}},}\
  }\href {\doibase 10.1103/PhysRevD.76.043510} {\bibfield  {journal} {\bibinfo
  {journal} {\prd}\ }\textbf {\bibinfo {volume} {76}},\ \bibinfo {eid} {043510}
  (\bibinfo {year} {2007}{\natexlab{a}})},\ \Eprint
  {http://arxiv.org/abs/0705.3980} {arXiv:0705.3980 [astro-ph]} \BibitemShut
  {NoStop}%
\bibitem [{\citenamefont {{Kitaura}}\ and\ \citenamefont
  {{En{\ss}lin}}(2008)}]{2008MNRAS.389..497K}%
  \BibitemOpen
  \bibfield  {author} {\bibinfo {author} {\bibfnamefont {F.~S.}\ \bibnamefont
  {{Kitaura}}}\ and\ \bibinfo {author} {\bibfnamefont {T.~A.}\ \bibnamefont
  {{En{\ss}lin}}},\ }\bibfield  {title} {\enquote {\bibinfo {title} {{Bayesian
  reconstruction of the cosmological large-scale structure: methodology,
  inverse algorithms and numerical optimization}},}\ }\href {\doibase
  10.1111/j.1365-2966.2008.13341.x} {\bibfield  {journal} {\bibinfo  {journal}
  {\mnras}\ }\textbf {\bibinfo {volume} {389}},\ \bibinfo {pages} {497--544}
  (\bibinfo {year} {2008})},\ \Eprint {http://arxiv.org/abs/0705.0429}
  {arXiv:0705.0429 [astro-ph]} \BibitemShut {NoStop}%
\bibitem [{\citenamefont {{Seljak}}\ \emph {et~al.}(2017)\citenamefont
  {{Seljak}}, \citenamefont {{Aslanyan}}, \citenamefont {{Feng}},\ and\
  \citenamefont {{Modi}}}]{2017JCAP...12..009S}%
  \BibitemOpen
  \bibfield  {author} {\bibinfo {author} {\bibfnamefont {Uro{\v{s}}}\
  \bibnamefont {{Seljak}}}, \bibinfo {author} {\bibfnamefont {Grigor}\
  \bibnamefont {{Aslanyan}}}, \bibinfo {author} {\bibfnamefont
  {Yu}~\bibnamefont {{Feng}}}, \ and\ \bibinfo {author} {\bibfnamefont
  {Chirag}\ \bibnamefont {{Modi}}},\ }\bibfield  {title} {\enquote {\bibinfo
  {title} {{Towards optimal extraction of cosmological information from
  nonlinear data}},}\ }\href {\doibase 10.1088/1475-7516/2017/12/009}
  {\bibfield  {journal} {\bibinfo  {journal} {Journal of Cosmology and
  Astro-Particle Physics}\ }\textbf {\bibinfo {volume} {2017}},\ \bibinfo {eid}
  {009} (\bibinfo {year} {2017})}\BibitemShut {NoStop}%
\bibitem [{\citenamefont {{Font-Ribera}}\ \emph {et~al.}(2018)\citenamefont
  {{Font-Ribera}}, \citenamefont {{McDonald}},\ and\ \citenamefont
  {{Slosar}}}]{2018JCAP...01..003F}%
  \BibitemOpen
  \bibfield  {author} {\bibinfo {author} {\bibfnamefont {A.}~\bibnamefont
  {{Font-Ribera}}}, \bibinfo {author} {\bibfnamefont {P.}~\bibnamefont
  {{McDonald}}}, \ and\ \bibinfo {author} {\bibfnamefont {A.}~\bibnamefont
  {{Slosar}}},\ }\bibfield  {title} {\enquote {\bibinfo {title} {{How to
  estimate the 3D power spectrum of the Lyman-{$\alpha$} forest}},}\ }\href
  {\doibase 10.1088/1475-7516/2018/01/003} {\bibfield  {journal} {\bibinfo
  {journal} {\jcap}\ }\textbf {\bibinfo {volume} {1}},\ \bibinfo {eid} {003}
  (\bibinfo {year} {2018})},\ \Eprint {http://arxiv.org/abs/1710.11036}
  {arXiv:1710.11036} \BibitemShut {NoStop}%
\bibitem [{\citenamefont {{Aghanim}}\ \emph {et~al.}(2016)\citenamefont
  {{Aghanim}}, \citenamefont {{Arnaud}}, \citenamefont {{Ashdown}},
  \citenamefont {{Aumont}}, \citenamefont {{Baccigalupi}}, \citenamefont
  {{Banday}}, \citenamefont {{Barreiro}}, \citenamefont {{Bartlett}},
  \citenamefont {{Bartolo}}, \citenamefont {{Battaner}}, \citenamefont
  {{Benabed}}, \citenamefont {{Beno{\^\i}t}}, \citenamefont
  {{Benoit-L{\'e}vy}}, \citenamefont {{Bernard}}, \citenamefont {{Bersanelli}},
  \citenamefont {{Bielewicz}}, \citenamefont {{Bock}}, \citenamefont
  {{Bonaldi}}, \citenamefont {{Bonavera}}, \citenamefont {{Bond}},
  \citenamefont {{Borrill}}, \citenamefont {{Bouchet}}, \citenamefont
  {{Boulanger}}, \citenamefont {{Bucher}}, \citenamefont {{Burigana}},
  \citenamefont {{Butler}}, \citenamefont {{Calabrese}}, \citenamefont
  {{Cardoso}}, \citenamefont {{Catalano}}, \citenamefont {{Challinor}},
  \citenamefont {{Chiang}}, \citenamefont {{Christensen}}, \citenamefont
  {{Clements}}, \citenamefont {{Colombo}}, \citenamefont {{Combet}},
  \citenamefont {{Coulais}}, \citenamefont {{Crill}}, \citenamefont {{Curto}},
  \citenamefont {{Cuttaia}}, \citenamefont {{Danese}}, \citenamefont
  {{Davies}}, \citenamefont {{Davis}}, \citenamefont {{de Bernardis}},
  \citenamefont {{de Rosa}}, \citenamefont {{de Zotti}}, \citenamefont
  {{Delabrouille}}, \citenamefont {{D{\'e}sert}}, \citenamefont {{Di
  Valentino}}, \citenamefont {{Dickinson}}, \citenamefont {{Diego}},
  \citenamefont {{Dolag}}, \citenamefont {{Dole}}, \citenamefont {{Donzelli}},
  \citenamefont {{Dor{\'e}}}, \citenamefont {{Douspis}}, \citenamefont
  {{Ducout}}, \citenamefont {{Dunkley}}, \citenamefont {{Dupac}}, \citenamefont
  {{Efstathiou}}, \citenamefont {{Elsner}}, \citenamefont {{En{\ss}lin}},
  \citenamefont {{Eriksen}}, \citenamefont {{Fergusson}}, \citenamefont
  {{Finelli}}, \citenamefont {{Forni}}, \citenamefont {{Frailis}},
  \citenamefont {{Fraisse}}, \citenamefont {{Franceschi}}, \citenamefont
  {{Frejsel}}, \citenamefont {{Galeotta}}, \citenamefont {{Galli}},
  \citenamefont {{Ganga}}, \citenamefont {{Gauthier}}, \citenamefont
  {{Gerbino}}, \citenamefont {{Giard}}, \citenamefont {{Gjerl{\o}w}},
  \citenamefont {{Gonz{\'a }lez-Nuevo}}, \citenamefont {{G{\'o}rski}},
  \citenamefont {{Gratton}}, \citenamefont {{Gregorio}}, \citenamefont
  {{Gruppuso}}, \citenamefont {{Gudmundsson}}, \citenamefont {{Hamann}},
  \citenamefont {{Hansen}}, \citenamefont {{Harrison}}, \citenamefont
  {{Helou}}, \citenamefont {{Henrot-Versill{\'e}}}, \citenamefont
  {{Hern{\'a}ndez- Monteagudo}}, \citenamefont {{Herranz}}, \citenamefont
  {{Hildebrandt}}, \citenamefont {{Hivon}}, \citenamefont {{Holmes}},
  \citenamefont {{Hornstrup}}, \citenamefont {{Huffenberger}}, \citenamefont
  {{Hurier}}, \citenamefont {{Jaffe}}, \citenamefont {{Jones}}, \citenamefont
  {{Juvela}}, \citenamefont {{Keih{\"a}nen}}, \citenamefont {{Keskitalo}},
  \citenamefont {{Kiiveri}}, \citenamefont {{Knoche}}, \citenamefont {{Knox}},
  \citenamefont {{Kunz}}, \citenamefont {{Kurki-Suonio}}, \citenamefont
  {{Lagache}}, \citenamefont {{L{\"a}hteenm{\"a}ki}}, \citenamefont
  {{Lamarre}}, \citenamefont {{Lasenby}}, \citenamefont {{Lattanzi}},
  \citenamefont {{Lawrence}}, \citenamefont {{Le Jeune}}, \citenamefont
  {{Leonardi}}, \citenamefont {{Lesgourgues}}, \citenamefont {{Levrier}},
  \citenamefont {{Lewis}}, \citenamefont {{Liguori}}, \citenamefont {{Lilje}},
  \citenamefont {{Lilley}}, \citenamefont {{Linden-V{\o}rnle}}, \citenamefont
  {{Lindholm}}, \citenamefont {{L{\'o}pez- Caniego}}, \citenamefont
  {{Mac{\'\i}as-P{\'e}rez}}, \citenamefont {{Maffei}}, \citenamefont
  {{Maggio}}, \citenamefont {{Maino}}, \citenamefont {{Mandolesi}},
  \citenamefont {{Mangilli}}, \citenamefont {{Maris}}, \citenamefont
  {{Martin}}, \citenamefont {{Mart{\'\i}nez-Gonz{\'a}lez}}, \citenamefont
  {{Masi}}, \citenamefont {{Matarrese}}, \citenamefont {{Meinhold}},
  \citenamefont {{Melchiorri}}, \citenamefont {{Migliaccio}}, \citenamefont
  {{Millea}}, \citenamefont {{Mitra}}, \citenamefont {{Miville-Desch{\^e}nes}},
  \citenamefont {{Moneti}}, \citenamefont {{Montier}}, \citenamefont
  {{Morgante}}, \citenamefont {{Mortlock}}, \citenamefont {{Mottet}},
  \citenamefont {{Munshi}}, \citenamefont {{Murphy}}, \citenamefont
  {{Narimani}}, \citenamefont {{Naselsky}}, \citenamefont {{Nati}},
  \citenamefont {{Natoli}}, \citenamefont {{Noviello}}, \citenamefont
  {{Novikov}}, \citenamefont {{Novikov}}, \citenamefont {{Oxborrow}},
  \citenamefont {{Paci}}, \citenamefont {{Pagano}}, \citenamefont {{Pajot}},
  \citenamefont {{Paoletti}}, \citenamefont {{Partridge}}, \citenamefont
  {{Pasian}}, \citenamefont {{Patanchon}}, \citenamefont {{Pearson}},
  \citenamefont {{Perdereau}}, \citenamefont {{Perotto}}, \citenamefont
  {{Pettorino}}, \citenamefont {{Piacentini}}, \citenamefont {{Piat}},
  \citenamefont {{Pierpaoli}}, \citenamefont {{Pietrobon}}, \citenamefont
  {{Plaszczynski}}, \citenamefont {{Pointecouteau}}, \citenamefont {{Polenta}},
  \citenamefont {{Ponthieu}}, \citenamefont {{Pratt}}, \citenamefont
  {{Prunet}}, \citenamefont {{Puget}}, \citenamefont {{Rachen}}, \citenamefont
  {{Reinecke}}, \citenamefont {{Remazeilles}}, \citenamefont {{Renault}},
  \citenamefont {{Renzi}}, \citenamefont {{Ristorcelli}}, \citenamefont
  {{Rocha}}, \citenamefont {{Rossetti}}, \citenamefont {{Roudier}},
  \citenamefont {{Rouill{\'e} d'Orfeuil}}, \citenamefont
  {{Rubi{\~n}o-Mart{\'\i}n}}, \citenamefont {{Rusholme}}, \citenamefont
  {{Salvati}}, \citenamefont {{Sandri}}, \citenamefont {{Santos}},
  \citenamefont {{Savelainen}}, \citenamefont {{Savini}}, \citenamefont
  {{Scott}}, \citenamefont {{Serra}}, \citenamefont {{Spencer}}, \citenamefont
  {{Spinelli}}, \citenamefont {{Stolyarov}}, \citenamefont {{Stompor}},
  \citenamefont {{Sunyaev}}, \citenamefont {{Sutton}}, \citenamefont
  {{Suur-Uski}}, \citenamefont {{Sygnet}}, \citenamefont {{Tauber}},
  \citenamefont {{Terenzi}}, \citenamefont {{Toffolatti}}, \citenamefont
  {{Tomasi}}, \citenamefont {{Tristram}}, \citenamefont {{Trombetti}},
  \citenamefont {{Tucci}}, \citenamefont {{Tuovinen}}, \citenamefont {{Umana}},
  \citenamefont {{Valenziano}}, \citenamefont {{Valiviita}}, \citenamefont
  {{Van Tent}}, \citenamefont {{Vielva}}, \citenamefont {{Villa}},
  \citenamefont {{Wade}}, \citenamefont {{Wandelt}}, \citenamefont {{Wehus}},
  \citenamefont {{Yvon}}, \citenamefont {{Zacchei}},\ and\ \citenamefont
  {{Zonca}}}]{2016A&A...594A..11P}%
  \BibitemOpen
  \bibfield  {author} {\bibinfo {author} {\bibfnamefont {N.}~\bibnamefont
  {{Aghanim}}}, \bibinfo {author} {\bibfnamefont {M.}~\bibnamefont {{Arnaud}}},
  \bibinfo {author} {\bibfnamefont {M.}~\bibnamefont {{Ashdown}}}, \bibinfo
  {author} {\bibfnamefont {J.}~\bibnamefont {{Aumont}}}, \bibinfo {author}
  {\bibfnamefont {C.}~\bibnamefont {{Baccigalupi}}}, \bibinfo {author}
  {\bibfnamefont {A.~J.}\ \bibnamefont {{Banday}}}, \bibinfo {author}
  {\bibfnamefont {R.~B.}\ \bibnamefont {{Barreiro}}}, \bibinfo {author}
  {\bibfnamefont {J.~G.}\ \bibnamefont {{Bartlett}}}, \bibinfo {author}
  {\bibfnamefont {N.}~\bibnamefont {{Bartolo}}}, \bibinfo {author}
  {\bibfnamefont {E.}~\bibnamefont {{Battaner}}}, \bibinfo {author}
  {\bibfnamefont {K.}~\bibnamefont {{Benabed}}}, \bibinfo {author}
  {\bibfnamefont {A.}~\bibnamefont {{Beno{\^\i}t}}}, \bibinfo {author}
  {\bibfnamefont {A.}~\bibnamefont {{Benoit-L{\'e}vy}}}, \bibinfo {author}
  {\bibfnamefont {J.~P.}\ \bibnamefont {{Bernard}}}, \bibinfo {author}
  {\bibfnamefont {M.}~\bibnamefont {{Bersanelli}}}, \bibinfo {author}
  {\bibfnamefont {P.}~\bibnamefont {{Bielewicz}}}, \bibinfo {author}
  {\bibfnamefont {J.~J.}\ \bibnamefont {{Bock}}}, \bibinfo {author}
  {\bibfnamefont {A.}~\bibnamefont {{Bonaldi}}}, \bibinfo {author}
  {\bibfnamefont {L.}~\bibnamefont {{Bonavera}}}, \bibinfo {author}
  {\bibfnamefont {J.~R.}\ \bibnamefont {{Bond}}}, \bibinfo {author}
  {\bibfnamefont {J.}~\bibnamefont {{Borrill}}}, \bibinfo {author}
  {\bibfnamefont {F.~R.}\ \bibnamefont {{Bouchet}}}, \bibinfo {author}
  {\bibfnamefont {F.}~\bibnamefont {{Boulanger}}}, \bibinfo {author}
  {\bibfnamefont {M.}~\bibnamefont {{Bucher}}}, \bibinfo {author}
  {\bibfnamefont {C.}~\bibnamefont {{Burigana}}}, \bibinfo {author}
  {\bibfnamefont {R.~C.}\ \bibnamefont {{Butler}}}, \bibinfo {author}
  {\bibfnamefont {E.}~\bibnamefont {{Calabrese}}}, \bibinfo {author}
  {\bibfnamefont {J.~F.}\ \bibnamefont {{Cardoso}}}, \bibinfo {author}
  {\bibfnamefont {A.}~\bibnamefont {{Catalano}}}, \bibinfo {author}
  {\bibfnamefont {A.}~\bibnamefont {{Challinor}}}, \bibinfo {author}
  {\bibfnamefont {H.~C.}\ \bibnamefont {{Chiang}}}, \bibinfo {author}
  {\bibfnamefont {P.~R.}\ \bibnamefont {{Christensen}}}, \bibinfo {author}
  {\bibfnamefont {D.~L.}\ \bibnamefont {{Clements}}}, \bibinfo {author}
  {\bibfnamefont {L.~P.~L.}\ \bibnamefont {{Colombo}}}, \bibinfo {author}
  {\bibfnamefont {C.}~\bibnamefont {{Combet}}}, \bibinfo {author}
  {\bibfnamefont {A.}~\bibnamefont {{Coulais}}}, \bibinfo {author}
  {\bibfnamefont {B.~P.}\ \bibnamefont {{Crill}}}, \bibinfo {author}
  {\bibfnamefont {A.}~\bibnamefont {{Curto}}}, \bibinfo {author} {\bibfnamefont
  {F.}~\bibnamefont {{Cuttaia}}}, \bibinfo {author} {\bibfnamefont
  {L.}~\bibnamefont {{Danese}}}, \bibinfo {author} {\bibfnamefont {R.~D.}\
  \bibnamefont {{Davies}}}, \bibinfo {author} {\bibfnamefont {R.~J.}\
  \bibnamefont {{Davis}}}, \bibinfo {author} {\bibfnamefont {P.}~\bibnamefont
  {{de Bernardis}}}, \bibinfo {author} {\bibfnamefont {A.}~\bibnamefont {{de
  Rosa}}}, \bibinfo {author} {\bibfnamefont {G.}~\bibnamefont {{de Zotti}}},
  \bibinfo {author} {\bibfnamefont {J.}~\bibnamefont {{Delabrouille}}},
  \bibinfo {author} {\bibfnamefont {F.~X.}\ \bibnamefont {{D{\'e}sert}}},
  \bibinfo {author} {\bibfnamefont {E.}~\bibnamefont {{Di Valentino}}},
  \bibinfo {author} {\bibfnamefont {C.}~\bibnamefont {{Dickinson}}}, \bibinfo
  {author} {\bibfnamefont {J.~M.}\ \bibnamefont {{Diego}}}, \bibinfo {author}
  {\bibfnamefont {K.}~\bibnamefont {{Dolag}}}, \bibinfo {author} {\bibfnamefont
  {H.}~\bibnamefont {{Dole}}}, \bibinfo {author} {\bibfnamefont
  {S.}~\bibnamefont {{Donzelli}}}, \bibinfo {author} {\bibfnamefont
  {O.}~\bibnamefont {{Dor{\'e}}}}, \bibinfo {author} {\bibfnamefont
  {M.}~\bibnamefont {{Douspis}}}, \bibinfo {author} {\bibfnamefont
  {A.}~\bibnamefont {{Ducout}}}, \bibinfo {author} {\bibfnamefont
  {J.}~\bibnamefont {{Dunkley}}}, \bibinfo {author} {\bibfnamefont
  {X.}~\bibnamefont {{Dupac}}}, \bibinfo {author} {\bibfnamefont
  {G.}~\bibnamefont {{Efstathiou}}}, \bibinfo {author} {\bibfnamefont
  {F.}~\bibnamefont {{Elsner}}}, \bibinfo {author} {\bibfnamefont {T.~A.}\
  \bibnamefont {{En{\ss}lin}}}, \bibinfo {author} {\bibfnamefont {H.~K.}\
  \bibnamefont {{Eriksen}}}, \bibinfo {author} {\bibfnamefont {J.}~\bibnamefont
  {{Fergusson}}}, \bibinfo {author} {\bibfnamefont {F.}~\bibnamefont
  {{Finelli}}}, \bibinfo {author} {\bibfnamefont {O.}~\bibnamefont {{Forni}}},
  \bibinfo {author} {\bibfnamefont {M.}~\bibnamefont {{Frailis}}}, \bibinfo
  {author} {\bibfnamefont {A.~A.}\ \bibnamefont {{Fraisse}}}, \bibinfo {author}
  {\bibfnamefont {E.}~\bibnamefont {{Franceschi}}}, \bibinfo {author}
  {\bibfnamefont {A.}~\bibnamefont {{Frejsel}}}, \bibinfo {author}
  {\bibfnamefont {S.}~\bibnamefont {{Galeotta}}}, \bibinfo {author}
  {\bibfnamefont {S.}~\bibnamefont {{Galli}}}, \bibinfo {author} {\bibfnamefont
  {K.}~\bibnamefont {{Ganga}}}, \bibinfo {author} {\bibfnamefont
  {C.}~\bibnamefont {{Gauthier}}}, \bibinfo {author} {\bibfnamefont
  {M.}~\bibnamefont {{Gerbino}}}, \bibinfo {author} {\bibfnamefont
  {M.}~\bibnamefont {{Giard}}}, \bibinfo {author} {\bibfnamefont
  {E.}~\bibnamefont {{Gjerl{\o}w}}}, \bibinfo {author} {\bibfnamefont
  {J.}~\bibnamefont {{Gonz{\'a }lez-Nuevo}}}, \bibinfo {author} {\bibfnamefont
  {K.~M.}\ \bibnamefont {{G{\'o}rski}}}, \bibinfo {author} {\bibfnamefont
  {S.}~\bibnamefont {{Gratton}}}, \bibinfo {author} {\bibfnamefont
  {A.}~\bibnamefont {{Gregorio}}}, \bibinfo {author} {\bibfnamefont
  {A.}~\bibnamefont {{Gruppuso}}}, \bibinfo {author} {\bibfnamefont {J.~E.}\
  \bibnamefont {{Gudmundsson}}}, \bibinfo {author} {\bibfnamefont
  {J.}~\bibnamefont {{Hamann}}}, \bibinfo {author} {\bibfnamefont {F.~K.}\
  \bibnamefont {{Hansen}}}, \bibinfo {author} {\bibfnamefont {D.~L.}\
  \bibnamefont {{Harrison}}}, \bibinfo {author} {\bibfnamefont
  {G.}~\bibnamefont {{Helou}}}, \bibinfo {author} {\bibfnamefont
  {S.}~\bibnamefont {{Henrot-Versill{\'e}}}}, \bibinfo {author} {\bibfnamefont
  {C.}~\bibnamefont {{Hern{\'a}ndez- Monteagudo}}}, \bibinfo {author}
  {\bibfnamefont {D.}~\bibnamefont {{Herranz}}}, \bibinfo {author}
  {\bibfnamefont {S.~R.}\ \bibnamefont {{Hildebrandt}}}, \bibinfo {author}
  {\bibfnamefont {E.}~\bibnamefont {{Hivon}}}, \bibinfo {author} {\bibfnamefont
  {W.~A.}\ \bibnamefont {{Holmes}}}, \bibinfo {author} {\bibfnamefont
  {A.}~\bibnamefont {{Hornstrup}}}, \bibinfo {author} {\bibfnamefont {K.~M.}\
  \bibnamefont {{Huffenberger}}}, \bibinfo {author} {\bibfnamefont
  {G.}~\bibnamefont {{Hurier}}}, \bibinfo {author} {\bibfnamefont {A.~H.}\
  \bibnamefont {{Jaffe}}}, \bibinfo {author} {\bibfnamefont {W.~C.}\
  \bibnamefont {{Jones}}}, \bibinfo {author} {\bibfnamefont {M.}~\bibnamefont
  {{Juvela}}}, \bibinfo {author} {\bibfnamefont {E.}~\bibnamefont
  {{Keih{\"a}nen}}}, \bibinfo {author} {\bibfnamefont {R.}~\bibnamefont
  {{Keskitalo}}}, \bibinfo {author} {\bibfnamefont {K.}~\bibnamefont
  {{Kiiveri}}}, \bibinfo {author} {\bibfnamefont {J.}~\bibnamefont {{Knoche}}},
  \bibinfo {author} {\bibfnamefont {L.}~\bibnamefont {{Knox}}}, \bibinfo
  {author} {\bibfnamefont {M.}~\bibnamefont {{Kunz}}}, \bibinfo {author}
  {\bibfnamefont {H.}~\bibnamefont {{Kurki-Suonio}}}, \bibinfo {author}
  {\bibfnamefont {G.}~\bibnamefont {{Lagache}}}, \bibinfo {author}
  {\bibfnamefont {A.}~\bibnamefont {{L{\"a}hteenm{\"a}ki}}}, \bibinfo {author}
  {\bibfnamefont {J.~M.}\ \bibnamefont {{Lamarre}}}, \bibinfo {author}
  {\bibfnamefont {A.}~\bibnamefont {{Lasenby}}}, \bibinfo {author}
  {\bibfnamefont {M.}~\bibnamefont {{Lattanzi}}}, \bibinfo {author}
  {\bibfnamefont {C.~R.}\ \bibnamefont {{Lawrence}}}, \bibinfo {author}
  {\bibfnamefont {M.}~\bibnamefont {{Le Jeune}}}, \bibinfo {author}
  {\bibfnamefont {R.}~\bibnamefont {{Leonardi}}}, \bibinfo {author}
  {\bibfnamefont {J.}~\bibnamefont {{Lesgourgues}}}, \bibinfo {author}
  {\bibfnamefont {F.}~\bibnamefont {{Levrier}}}, \bibinfo {author}
  {\bibfnamefont {A.}~\bibnamefont {{Lewis}}}, \bibinfo {author} {\bibfnamefont
  {M.}~\bibnamefont {{Liguori}}}, \bibinfo {author} {\bibfnamefont {P.~B.}\
  \bibnamefont {{Lilje}}}, \bibinfo {author} {\bibfnamefont {M.}~\bibnamefont
  {{Lilley}}}, \bibinfo {author} {\bibfnamefont {M.}~\bibnamefont
  {{Linden-V{\o}rnle}}}, \bibinfo {author} {\bibfnamefont {V.}~\bibnamefont
  {{Lindholm}}}, \bibinfo {author} {\bibfnamefont {M.}~\bibnamefont
  {{L{\'o}pez- Caniego}}}, \bibinfo {author} {\bibfnamefont {J.~F.}\
  \bibnamefont {{Mac{\'\i}as-P{\'e}rez}}}, \bibinfo {author} {\bibfnamefont
  {B.}~\bibnamefont {{Maffei}}}, \bibinfo {author} {\bibfnamefont
  {G.}~\bibnamefont {{Maggio}}}, \bibinfo {author} {\bibfnamefont
  {D.}~\bibnamefont {{Maino}}}, \bibinfo {author} {\bibfnamefont
  {N.}~\bibnamefont {{Mandolesi}}}, \bibinfo {author} {\bibfnamefont
  {A.}~\bibnamefont {{Mangilli}}}, \bibinfo {author} {\bibfnamefont
  {M.}~\bibnamefont {{Maris}}}, \bibinfo {author} {\bibfnamefont {P.~G.}\
  \bibnamefont {{Martin}}}, \bibinfo {author} {\bibfnamefont {E.}~\bibnamefont
  {{Mart{\'\i}nez-Gonz{\'a}lez}}}, \bibinfo {author} {\bibfnamefont
  {S.}~\bibnamefont {{Masi}}}, \bibinfo {author} {\bibfnamefont
  {S.}~\bibnamefont {{Matarrese}}}, \bibinfo {author} {\bibfnamefont {P.~R.}\
  \bibnamefont {{Meinhold}}}, \bibinfo {author} {\bibfnamefont
  {A.}~\bibnamefont {{Melchiorri}}}, \bibinfo {author} {\bibfnamefont
  {M.}~\bibnamefont {{Migliaccio}}}, \bibinfo {author} {\bibfnamefont
  {M.}~\bibnamefont {{Millea}}}, \bibinfo {author} {\bibfnamefont
  {S.}~\bibnamefont {{Mitra}}}, \bibinfo {author} {\bibfnamefont {M.~A.}\
  \bibnamefont {{Miville-Desch{\^e}nes}}}, \bibinfo {author} {\bibfnamefont
  {A.}~\bibnamefont {{Moneti}}}, \bibinfo {author} {\bibfnamefont
  {L.}~\bibnamefont {{Montier}}}, \bibinfo {author} {\bibfnamefont
  {G.}~\bibnamefont {{Morgante}}}, \bibinfo {author} {\bibfnamefont
  {D.}~\bibnamefont {{Mortlock}}}, \bibinfo {author} {\bibfnamefont
  {S.}~\bibnamefont {{Mottet}}}, \bibinfo {author} {\bibfnamefont
  {D.}~\bibnamefont {{Munshi}}}, \bibinfo {author} {\bibfnamefont {J.~A.}\
  \bibnamefont {{Murphy}}}, \bibinfo {author} {\bibfnamefont {A.}~\bibnamefont
  {{Narimani}}}, \bibinfo {author} {\bibfnamefont {P.}~\bibnamefont
  {{Naselsky}}}, \bibinfo {author} {\bibfnamefont {F.}~\bibnamefont {{Nati}}},
  \bibinfo {author} {\bibfnamefont {P.}~\bibnamefont {{Natoli}}}, \bibinfo
  {author} {\bibfnamefont {F.}~\bibnamefont {{Noviello}}}, \bibinfo {author}
  {\bibfnamefont {D.}~\bibnamefont {{Novikov}}}, \bibinfo {author}
  {\bibfnamefont {I.}~\bibnamefont {{Novikov}}}, \bibinfo {author}
  {\bibfnamefont {C.~A.}\ \bibnamefont {{Oxborrow}}}, \bibinfo {author}
  {\bibfnamefont {F.}~\bibnamefont {{Paci}}}, \bibinfo {author} {\bibfnamefont
  {L.}~\bibnamefont {{Pagano}}}, \bibinfo {author} {\bibfnamefont
  {F.}~\bibnamefont {{Pajot}}}, \bibinfo {author} {\bibfnamefont
  {D.}~\bibnamefont {{Paoletti}}}, \bibinfo {author} {\bibfnamefont
  {B.}~\bibnamefont {{Partridge}}}, \bibinfo {author} {\bibfnamefont
  {F.}~\bibnamefont {{Pasian}}}, \bibinfo {author} {\bibfnamefont
  {G.}~\bibnamefont {{Patanchon}}}, \bibinfo {author} {\bibfnamefont {T.~J.}\
  \bibnamefont {{Pearson}}}, \bibinfo {author} {\bibfnamefont {O.}~\bibnamefont
  {{Perdereau}}}, \bibinfo {author} {\bibfnamefont {L.}~\bibnamefont
  {{Perotto}}}, \bibinfo {author} {\bibfnamefont {V.}~\bibnamefont
  {{Pettorino}}}, \bibinfo {author} {\bibfnamefont {F.}~\bibnamefont
  {{Piacentini}}}, \bibinfo {author} {\bibfnamefont {M.}~\bibnamefont
  {{Piat}}}, \bibinfo {author} {\bibfnamefont {E.}~\bibnamefont {{Pierpaoli}}},
  \bibinfo {author} {\bibfnamefont {D.}~\bibnamefont {{Pietrobon}}}, \bibinfo
  {author} {\bibfnamefont {S.}~\bibnamefont {{Plaszczynski}}}, \bibinfo
  {author} {\bibfnamefont {E.}~\bibnamefont {{Pointecouteau}}}, \bibinfo
  {author} {\bibfnamefont {G.}~\bibnamefont {{Polenta}}}, \bibinfo {author}
  {\bibfnamefont {N.}~\bibnamefont {{Ponthieu}}}, \bibinfo {author}
  {\bibfnamefont {G.~W.}\ \bibnamefont {{Pratt}}}, \bibinfo {author}
  {\bibfnamefont {S.}~\bibnamefont {{Prunet}}}, \bibinfo {author}
  {\bibfnamefont {J.~L.}\ \bibnamefont {{Puget}}}, \bibinfo {author}
  {\bibfnamefont {J.~P.}\ \bibnamefont {{Rachen}}}, \bibinfo {author}
  {\bibfnamefont {M.}~\bibnamefont {{Reinecke}}}, \bibinfo {author}
  {\bibfnamefont {M.}~\bibnamefont {{Remazeilles}}}, \bibinfo {author}
  {\bibfnamefont {C.}~\bibnamefont {{Renault}}}, \bibinfo {author}
  {\bibfnamefont {A.}~\bibnamefont {{Renzi}}}, \bibinfo {author} {\bibfnamefont
  {I.}~\bibnamefont {{Ristorcelli}}}, \bibinfo {author} {\bibfnamefont
  {G.}~\bibnamefont {{Rocha}}}, \bibinfo {author} {\bibfnamefont
  {M.}~\bibnamefont {{Rossetti}}}, \bibinfo {author} {\bibfnamefont
  {G.}~\bibnamefont {{Roudier}}}, \bibinfo {author} {\bibfnamefont
  {B.}~\bibnamefont {{Rouill{\'e} d'Orfeuil}}}, \bibinfo {author}
  {\bibfnamefont {J.~A.}\ \bibnamefont {{Rubi{\~n}o-Mart{\'\i}n}}}, \bibinfo
  {author} {\bibfnamefont {B.}~\bibnamefont {{Rusholme}}}, \bibinfo {author}
  {\bibfnamefont {L.}~\bibnamefont {{Salvati}}}, \bibinfo {author}
  {\bibfnamefont {M.}~\bibnamefont {{Sandri}}}, \bibinfo {author}
  {\bibfnamefont {D.}~\bibnamefont {{Santos}}}, \bibinfo {author}
  {\bibfnamefont {M.}~\bibnamefont {{Savelainen}}}, \bibinfo {author}
  {\bibfnamefont {G.}~\bibnamefont {{Savini}}}, \bibinfo {author}
  {\bibfnamefont {D.}~\bibnamefont {{Scott}}}, \bibinfo {author} {\bibfnamefont
  {P.}~\bibnamefont {{Serra}}}, \bibinfo {author} {\bibfnamefont {L.~D.}\
  \bibnamefont {{Spencer}}}, \bibinfo {author} {\bibfnamefont {M.}~\bibnamefont
  {{Spinelli}}}, \bibinfo {author} {\bibfnamefont {V.}~\bibnamefont
  {{Stolyarov}}}, \bibinfo {author} {\bibfnamefont {R.}~\bibnamefont
  {{Stompor}}}, \bibinfo {author} {\bibfnamefont {R.}~\bibnamefont
  {{Sunyaev}}}, \bibinfo {author} {\bibfnamefont {D.}~\bibnamefont {{Sutton}}},
  \bibinfo {author} {\bibfnamefont {A.~S.}\ \bibnamefont {{Suur-Uski}}},
  \bibinfo {author} {\bibfnamefont {J.~F.}\ \bibnamefont {{Sygnet}}}, \bibinfo
  {author} {\bibfnamefont {J.~A.}\ \bibnamefont {{Tauber}}}, \bibinfo {author}
  {\bibfnamefont {L.}~\bibnamefont {{Terenzi}}}, \bibinfo {author}
  {\bibfnamefont {L.}~\bibnamefont {{Toffolatti}}}, \bibinfo {author}
  {\bibfnamefont {M.}~\bibnamefont {{Tomasi}}}, \bibinfo {author}
  {\bibfnamefont {M.}~\bibnamefont {{Tristram}}}, \bibinfo {author}
  {\bibfnamefont {T.}~\bibnamefont {{Trombetti}}}, \bibinfo {author}
  {\bibfnamefont {M.}~\bibnamefont {{Tucci}}}, \bibinfo {author} {\bibfnamefont
  {J.}~\bibnamefont {{Tuovinen}}}, \bibinfo {author} {\bibfnamefont
  {G.}~\bibnamefont {{Umana}}}, \bibinfo {author} {\bibfnamefont
  {L.}~\bibnamefont {{Valenziano}}}, \bibinfo {author} {\bibfnamefont
  {J.}~\bibnamefont {{Valiviita}}}, \bibinfo {author} {\bibfnamefont
  {F.}~\bibnamefont {{Van Tent}}}, \bibinfo {author} {\bibfnamefont
  {P.}~\bibnamefont {{Vielva}}}, \bibinfo {author} {\bibfnamefont
  {F.}~\bibnamefont {{Villa}}}, \bibinfo {author} {\bibfnamefont {L.~A.}\
  \bibnamefont {{Wade}}}, \bibinfo {author} {\bibfnamefont {B.~D.}\
  \bibnamefont {{Wandelt}}}, \bibinfo {author} {\bibfnamefont {I.~K.}\
  \bibnamefont {{Wehus}}}, \bibinfo {author} {\bibfnamefont {D.}~\bibnamefont
  {{Yvon}}}, \bibinfo {author} {\bibfnamefont {A.}~\bibnamefont {{Zacchei}}}, \
  and\ \bibinfo {author} {\bibfnamefont {A.}~\bibnamefont {{Zonca}}},\
  }\bibfield  {title} {\enquote {\bibinfo {title} {{Planck 2015 results. XI.
  CMB power spectra, likelihoods, and robustness of parameters}},}\ }\href
  {\doibase 10.1051/0004-6361/201526926} {\bibfield  {journal} {\bibinfo
  {journal} {\aap}\ }\textbf {\bibinfo {volume} {594}},\ \bibinfo {eid} {A11}
  (\bibinfo {year} {2016})}\BibitemShut {NoStop}%
\bibitem [{\citenamefont {{Beutler}}\ \emph {et~al.}(2017)\citenamefont
  {{Beutler}}, \citenamefont {{Seo}}, \citenamefont {{Ross}}, \citenamefont
  {{McDonald}}, \citenamefont {{Saito}}, \citenamefont {{Bolton}},
  \citenamefont {{Brownstein}}, \citenamefont {{Chuang}}, \citenamefont
  {{Cuesta}}, \citenamefont {{Eisenstein}}, \citenamefont {{Font-Ribera}},
  \citenamefont {{Grieb}}, \citenamefont {{Hand}}, \citenamefont {{Kitaura}},
  \citenamefont {{Modi}}, \citenamefont {{Nichol}}, \citenamefont {{Percival}},
  \citenamefont {{Prada}}, \citenamefont {{Rodriguez-Torres}}, \citenamefont
  {{Roe}}, \citenamefont {{Ross}}, \citenamefont {{Salazar-Albornoz}},
  \citenamefont {{S{\'a}nchez}}, \citenamefont {{Schneider}}, \citenamefont
  {{Slosar}}, \citenamefont {{Tinker}}, \citenamefont {{Tojeiro}},
  \citenamefont {{Vargas-Maga{\~n}a}},\ and\ \citenamefont
  {{Vazquez}}}]{2017MNRAS.464.3409B}%
  \BibitemOpen
  \bibfield  {author} {\bibinfo {author} {\bibfnamefont {F.}~\bibnamefont
  {{Beutler}}}, \bibinfo {author} {\bibfnamefont {H.-J.}\ \bibnamefont
  {{Seo}}}, \bibinfo {author} {\bibfnamefont {A.~J.}\ \bibnamefont {{Ross}}},
  \bibinfo {author} {\bibfnamefont {P.}~\bibnamefont {{McDonald}}}, \bibinfo
  {author} {\bibfnamefont {S.}~\bibnamefont {{Saito}}}, \bibinfo {author}
  {\bibfnamefont {A.~S.}\ \bibnamefont {{Bolton}}}, \bibinfo {author}
  {\bibfnamefont {J.~R.}\ \bibnamefont {{Brownstein}}}, \bibinfo {author}
  {\bibfnamefont {C.-H.}\ \bibnamefont {{Chuang}}}, \bibinfo {author}
  {\bibfnamefont {A.~J.}\ \bibnamefont {{Cuesta}}}, \bibinfo {author}
  {\bibfnamefont {D.~J.}\ \bibnamefont {{Eisenstein}}}, \bibinfo {author}
  {\bibfnamefont {A.}~\bibnamefont {{Font-Ribera}}}, \bibinfo {author}
  {\bibfnamefont {J.~N.}\ \bibnamefont {{Grieb}}}, \bibinfo {author}
  {\bibfnamefont {N.}~\bibnamefont {{Hand}}}, \bibinfo {author} {\bibfnamefont
  {F.-S.}\ \bibnamefont {{Kitaura}}}, \bibinfo {author} {\bibfnamefont
  {C.}~\bibnamefont {{Modi}}}, \bibinfo {author} {\bibfnamefont {R.~C.}\
  \bibnamefont {{Nichol}}}, \bibinfo {author} {\bibfnamefont {W.~J.}\
  \bibnamefont {{Percival}}}, \bibinfo {author} {\bibfnamefont
  {F.}~\bibnamefont {{Prada}}}, \bibinfo {author} {\bibfnamefont
  {S.}~\bibnamefont {{Rodriguez-Torres}}}, \bibinfo {author} {\bibfnamefont
  {N.~A.}\ \bibnamefont {{Roe}}}, \bibinfo {author} {\bibfnamefont {N.~P.}\
  \bibnamefont {{Ross}}}, \bibinfo {author} {\bibfnamefont {S.}~\bibnamefont
  {{Salazar-Albornoz}}}, \bibinfo {author} {\bibfnamefont {A.~G.}\ \bibnamefont
  {{S{\'a}nchez}}}, \bibinfo {author} {\bibfnamefont {D.~P.}\ \bibnamefont
  {{Schneider}}}, \bibinfo {author} {\bibfnamefont {A.}~\bibnamefont
  {{Slosar}}}, \bibinfo {author} {\bibfnamefont {J.}~\bibnamefont {{Tinker}}},
  \bibinfo {author} {\bibfnamefont {R.}~\bibnamefont {{Tojeiro}}}, \bibinfo
  {author} {\bibfnamefont {M.}~\bibnamefont {{Vargas-Maga{\~n}a}}}, \ and\
  \bibinfo {author} {\bibfnamefont {J.~A.}\ \bibnamefont {{Vazquez}}},\
  }\bibfield  {title} {\enquote {\bibinfo {title} {{The clustering of galaxies
  in the completed SDSS-III Baryon Oscillation Spectroscopic Survey: baryon
  acoustic oscillations in the Fourier space}},}\ }\href {\doibase
  10.1093/mnras/stw2373} {\bibfield  {journal} {\bibinfo  {journal} {\mnras}\
  }\textbf {\bibinfo {volume} {464}},\ \bibinfo {pages} {3409--3430} (\bibinfo
  {year} {2017})},\ \Eprint {http://arxiv.org/abs/1607.03149}
  {arXiv:1607.03149} \BibitemShut {NoStop}%
\bibitem [{\citenamefont {{Wilson}}\ and\ \citenamefont
  {{Kogut}}(1974)}]{1974PhR....12...75W}%
  \BibitemOpen
  \bibfield  {author} {\bibinfo {author} {\bibfnamefont {K.~G.}\ \bibnamefont
  {{Wilson}}}\ and\ \bibinfo {author} {\bibfnamefont {J.}~\bibnamefont
  {{Kogut}}},\ }\bibfield  {title} {\enquote {\bibinfo {title} {{The
  renormalization group and the {$\epsilon$} expansion}},}\ }\href {\doibase
  10.1016/0370-1573(74)90023-4} {\bibfield  {journal} {\bibinfo  {journal}
  {\physrep}\ }\textbf {\bibinfo {volume} {12}},\ \bibinfo {pages} {75--199}
  (\bibinfo {year} {1974})}\BibitemShut {NoStop}%
\bibitem [{\citenamefont {{Banks}}(2008)}]{2008mqft.book.....B}%
  \BibitemOpen
  \bibfield  {author} {\bibinfo {author} {\bibfnamefont {T.}~\bibnamefont
  {{Banks}}},\ }\href@noop {} {\emph {\bibinfo {title} {Modern Quantum Field
  Theory, by Tom Banks, Cambridge, UK: Cambridge University Press, 2008}}}\
  (\bibinfo {year} {2008})\BibitemShut {NoStop}%
\bibitem [{Note1()}]{Note1}%
  \BibitemOpen
  \bibinfo {note} {One may wonder how generic is this behavior that differences
  in $\chi ^2$ between models have much better accuracy than the absolute value
  of $\chi ^2$. In practice it does not really matter because we will always
  test numerical convergence of the quantities of interest (generally
  differences between models) on a case-by-case basis, determining whether they
  converge without regard to whether they converge faster than absolute $\chi
  ^2$ or not, but maybe it is an interesting academic question. I think this
  behavior probably is generic, because of the usual relation between absolute
  $\chi ^2$, degrees of freedom (dof), and $\Delta \chi ^2$ between models.
  Absolute $\chi ^2$ for any data set has big random fluctuations around a big
  number dof, much more for a large data set than the level of $\Delta \chi ^2$
  we would call significant as a difference between models (e.g., we would say
  a model is ruled out decisively if it fits $\Delta \chi ^2=25$ worse than
  another one, even if both of them would look like they had a ``fine $\chi
  ^2$'', i.e., $\chi ^2\sim $ dof, if compared to that data in isolation). The
  reason these things work this way is that most of the contribution to
  absolute $\chi ^2$ is effectively orthogonal to any smooth model, so it is
  irrelevant to comparison between models. I think we can generically expect
  this likelihood calculation to behave similarly. Again though, we can/must
  always do convergence tests to verify that the likelihood differences that we
  care about are not sensitive to numerical parameters, so we won't ever be
  fooled by imperfection in this intuition.}\BibitemShut {Stop}%
\bibitem [{\citenamefont {{McDonald}}(2008)}]{2008PhRvD..78l3519M}%
  \BibitemOpen
  \bibfield  {author} {\bibinfo {author} {\bibfnamefont {P.}~\bibnamefont
  {{McDonald}}},\ }\bibfield  {title} {\enquote {\bibinfo {title} {{Primordial
  non-Gaussianity: Large-scale structure signature in the perturbative bias
  model}},}\ }\href {\doibase 10.1103/PhysRevD.78.123519} {\bibfield  {journal}
  {\bibinfo  {journal} {\prd}\ }\textbf {\bibinfo {volume} {78}},\ \bibinfo
  {pages} {123519--+} (\bibinfo {year} {2008})},\ \Eprint
  {http://arxiv.org/abs/0806.1061} {arXiv:0806.1061} \BibitemShut {NoStop}%
\bibitem [{\citenamefont {{Bartolo}}\ \emph {et~al.}(2010)\citenamefont
  {{Bartolo}}, \citenamefont {{Beltr{\'a}n Almeida}}, \citenamefont
  {{Matarrese}}, \citenamefont {{Pietroni}},\ and\ \citenamefont
  {{Riotto}}}]{2010JCAP...03..011B}%
  \BibitemOpen
  \bibfield  {author} {\bibinfo {author} {\bibfnamefont {N.}~\bibnamefont
  {{Bartolo}}}, \bibinfo {author} {\bibfnamefont {J.~P.}\ \bibnamefont
  {{Beltr{\'a}n Almeida}}}, \bibinfo {author} {\bibfnamefont {S.}~\bibnamefont
  {{Matarrese}}}, \bibinfo {author} {\bibfnamefont {M.}~\bibnamefont
  {{Pietroni}}}, \ and\ \bibinfo {author} {\bibfnamefont {A.}~\bibnamefont
  {{Riotto}}},\ }\bibfield  {title} {\enquote {\bibinfo {title} {{Signatures of
  primordial non-Gaussianities in the matter power-spectrum and bispectrum: the
  time-RG approach}},}\ }\href {\doibase 10.1088/1475-7516/2010/03/011}
  {\bibfield  {journal} {\bibinfo  {journal} {\jcap}\ }\textbf {\bibinfo
  {volume} {3}},\ \bibinfo {eid} {011} (\bibinfo {year} {2010})},\ \Eprint
  {http://arxiv.org/abs/0912.4276} {arXiv:0912.4276 [astro-ph.CO]} \BibitemShut
  {NoStop}%
\bibitem [{\citenamefont {{Giannantonio}}\ and\ \citenamefont
  {{Porciani}}(2010)}]{2010PhRvD..81f3530G}%
  \BibitemOpen
  \bibfield  {author} {\bibinfo {author} {\bibfnamefont {T.}~\bibnamefont
  {{Giannantonio}}}\ and\ \bibinfo {author} {\bibfnamefont {C.}~\bibnamefont
  {{Porciani}}},\ }\bibfield  {title} {\enquote {\bibinfo {title} {{Structure
  formation from non-Gaussian initial conditions: Multivariate biasing,
  statistics, and comparison with N-body simulations}},}\ }\href {\doibase
  10.1103/PhysRevD.81.063530} {\bibfield  {journal} {\bibinfo  {journal}
  {\prd}\ }\textbf {\bibinfo {volume} {81}},\ \bibinfo {eid} {063530} (\bibinfo
  {year} {2010})},\ \Eprint {http://arxiv.org/abs/0911.0017} {arXiv:0911.0017
  [astro-ph.CO]} \BibitemShut {NoStop}%
\bibitem [{\citenamefont {{Gong}}\ and\ \citenamefont
  {{Yokoyama}}(2011)}]{2011MNRAS.417L..79G}%
  \BibitemOpen
  \bibfield  {author} {\bibinfo {author} {\bibfnamefont {J.-O.}\ \bibnamefont
  {{Gong}}}\ and\ \bibinfo {author} {\bibfnamefont {S.}~\bibnamefont
  {{Yokoyama}}},\ }\bibfield  {title} {\enquote {\bibinfo {title}
  {{Scale-dependent bias from primordial non-Gaussianity with trispectrum}},}\
  }\href {\doibase 10.1111/j.1745-3933.2011.01124.x} {\bibfield  {journal}
  {\bibinfo  {journal} {\mnras}\ }\textbf {\bibinfo {volume} {417}},\ \bibinfo
  {pages} {L79--L82} (\bibinfo {year} {2011})},\ \Eprint
  {http://arxiv.org/abs/1106.4404} {arXiv:1106.4404 [astro-ph.CO]} \BibitemShut
  {NoStop}%
\bibitem [{\citenamefont {{Alvarez}}\ \emph {et~al.}(2014)\citenamefont
  {{Alvarez}}, \citenamefont {{Baldauf}}, \citenamefont {{Bond}}, \citenamefont
  {{Dalal}}, \citenamefont {{de Putter}}, \citenamefont {{Dor{\'e}}},
  \citenamefont {{Green}}, \citenamefont {{Hirata}}, \citenamefont {{Huang}},
  \citenamefont {{Huterer}}, \citenamefont {{Jeong}}, \citenamefont
  {{Johnson}}, \citenamefont {{Krause}}, \citenamefont {{Loverde}},
  \citenamefont {{Meyers}}, \citenamefont {{Meerburg}}, \citenamefont
  {{Senatore}}, \citenamefont {{Shandera}}, \citenamefont {{Silverstein}},
  \citenamefont {{Slosar}}, \citenamefont {{Smith}}, \citenamefont
  {{Zaldarriaga}}, \citenamefont {{Assassi}}, \citenamefont {{Braden}},
  \citenamefont {{Hajian}}, \citenamefont {{Kobayashi}}, \citenamefont
  {{Stein}},\ and\ \citenamefont {{van Engelen}}}]{2014arXiv1412.4671A}%
  \BibitemOpen
  \bibfield  {author} {\bibinfo {author} {\bibfnamefont {M.}~\bibnamefont
  {{Alvarez}}}, \bibinfo {author} {\bibfnamefont {T.}~\bibnamefont
  {{Baldauf}}}, \bibinfo {author} {\bibfnamefont {J.~R.}\ \bibnamefont
  {{Bond}}}, \bibinfo {author} {\bibfnamefont {N.}~\bibnamefont {{Dalal}}},
  \bibinfo {author} {\bibfnamefont {R.}~\bibnamefont {{de Putter}}}, \bibinfo
  {author} {\bibfnamefont {O.}~\bibnamefont {{Dor{\'e}}}}, \bibinfo {author}
  {\bibfnamefont {D.}~\bibnamefont {{Green}}}, \bibinfo {author} {\bibfnamefont
  {C.}~\bibnamefont {{Hirata}}}, \bibinfo {author} {\bibfnamefont
  {Z.}~\bibnamefont {{Huang}}}, \bibinfo {author} {\bibfnamefont
  {D.}~\bibnamefont {{Huterer}}}, \bibinfo {author} {\bibfnamefont
  {D.}~\bibnamefont {{Jeong}}}, \bibinfo {author} {\bibfnamefont {M.~C.}\
  \bibnamefont {{Johnson}}}, \bibinfo {author} {\bibfnamefont {E.}~\bibnamefont
  {{Krause}}}, \bibinfo {author} {\bibfnamefont {M.}~\bibnamefont {{Loverde}}},
  \bibinfo {author} {\bibfnamefont {J.}~\bibnamefont {{Meyers}}}, \bibinfo
  {author} {\bibfnamefont {P.~D.}\ \bibnamefont {{Meerburg}}}, \bibinfo
  {author} {\bibfnamefont {L.}~\bibnamefont {{Senatore}}}, \bibinfo {author}
  {\bibfnamefont {S.}~\bibnamefont {{Shandera}}}, \bibinfo {author}
  {\bibfnamefont {E.}~\bibnamefont {{Silverstein}}}, \bibinfo {author}
  {\bibfnamefont {A.}~\bibnamefont {{Slosar}}}, \bibinfo {author}
  {\bibfnamefont {K.}~\bibnamefont {{Smith}}}, \bibinfo {author} {\bibfnamefont
  {M.}~\bibnamefont {{Zaldarriaga}}}, \bibinfo {author} {\bibfnamefont
  {V.}~\bibnamefont {{Assassi}}}, \bibinfo {author} {\bibfnamefont
  {J.}~\bibnamefont {{Braden}}}, \bibinfo {author} {\bibfnamefont
  {A.}~\bibnamefont {{Hajian}}}, \bibinfo {author} {\bibfnamefont
  {T.}~\bibnamefont {{Kobayashi}}}, \bibinfo {author} {\bibfnamefont
  {G.}~\bibnamefont {{Stein}}}, \ and\ \bibinfo {author} {\bibfnamefont
  {A.}~\bibnamefont {{van Engelen}}},\ }\bibfield  {title} {\enquote {\bibinfo
  {title} {{Testing Inflation with Large Scale Structure: Connecting Hopes with
  Reality}},}\ }\href@noop {} {\bibfield  {journal} {\bibinfo  {journal} {ArXiv
  e-prints}\ } (\bibinfo {year} {2014})},\ \Eprint
  {http://arxiv.org/abs/1412.4671} {arXiv:1412.4671} \BibitemShut {NoStop}%
\bibitem [{\citenamefont {{Moradinezhad Dizgah}}\ and\ \citenamefont
  {{Dvorkin}}(2018)}]{2018JCAP...01..010M}%
  \BibitemOpen
  \bibfield  {author} {\bibinfo {author} {\bibfnamefont {Azadeh}\ \bibnamefont
  {{Moradinezhad Dizgah}}}\ and\ \bibinfo {author} {\bibfnamefont {Cora}\
  \bibnamefont {{Dvorkin}}},\ }\bibfield  {title} {\enquote {\bibinfo {title}
  {{Scale-dependent galaxy bias from massive particles with spin during
  inflation}},}\ }\href {\doibase 10.1088/1475-7516/2018/01/010} {\bibfield
  {journal} {\bibinfo  {journal} {Journal of Cosmology and Astro-Particle
  Physics}\ }\textbf {\bibinfo {volume} {2018}},\ \bibinfo {eid} {010}
  (\bibinfo {year} {2018})}\BibitemShut {NoStop}%
\bibitem [{\citenamefont {{Valageas}}(2001)}]{2001A&A...379....8V}%
  \BibitemOpen
  \bibfield  {author} {\bibinfo {author} {\bibfnamefont {P.}~\bibnamefont
  {{Valageas}}},\ }\bibfield  {title} {\enquote {\bibinfo {title} {{Dynamics of
  gravitational clustering. I. Building perturbative expansions}},}\ }\href
  {\doibase 10.1051/0004-6361:20011309} {\bibfield  {journal} {\bibinfo
  {journal} {\aap}\ }\textbf {\bibinfo {volume} {379}},\ \bibinfo {pages}
  {8--20} (\bibinfo {year} {2001})}\BibitemShut {NoStop}%
\bibitem [{\citenamefont {{Valageas}}(2002)}]{2002A&A...382..412V}%
  \BibitemOpen
  \bibfield  {author} {\bibinfo {author} {\bibfnamefont {P.}~\bibnamefont
  {{Valageas}}},\ }\bibfield  {title} {\enquote {\bibinfo {title} {{Dynamics of
  gravitational clustering. II. Steepest-descent method for the quasi-linear
  regime}},}\ }\href {\doibase 10.1051/0004-6361:20011663} {\bibfield
  {journal} {\bibinfo  {journal} {\aap}\ }\textbf {\bibinfo {volume} {382}},\
  \bibinfo {pages} {412--430} (\bibinfo {year} {2002})},\ \Eprint
  {http://arxiv.org/abs/astro-ph/0107126} {astro-ph/0107126} \BibitemShut
  {NoStop}%
\bibitem [{\citenamefont {{Valageas}}(2004)}]{2004A&A...421...23V}%
  \BibitemOpen
  \bibfield  {author} {\bibinfo {author} {\bibfnamefont {P.}~\bibnamefont
  {{Valageas}}},\ }\bibfield  {title} {\enquote {\bibinfo {title} {{A new
  approach to gravitational clustering: A path-integral formalism and large-N
  expansions}},}\ }\href {\doibase 10.1051/0004-6361:20040125} {\bibfield
  {journal} {\bibinfo  {journal} {\aap}\ }\textbf {\bibinfo {volume} {421}},\
  \bibinfo {pages} {23--40} (\bibinfo {year} {2004})},\ \Eprint
  {http://arxiv.org/abs/astro-ph/0307008} {astro-ph/0307008} \BibitemShut
  {NoStop}%
\bibitem [{\citenamefont {{Valageas}}(2007)}]{2007A&A...465..725V}%
  \BibitemOpen
  \bibfield  {author} {\bibinfo {author} {\bibfnamefont {P.}~\bibnamefont
  {{Valageas}}},\ }\bibfield  {title} {\enquote {\bibinfo {title} {{Large-N
  expansions applied to gravitational clustering}},}\ }\href {\doibase
  10.1051/0004-6361:20066832} {\bibfield  {journal} {\bibinfo  {journal}
  {\aap}\ }\textbf {\bibinfo {volume} {465}},\ \bibinfo {pages} {725--747}
  (\bibinfo {year} {2007})},\ \Eprint {http://arxiv.org/abs/astro-ph/0611849}
  {astro-ph/0611849} \BibitemShut {NoStop}%
\bibitem [{\citenamefont {{McDonald}}(2007)}]{2007PhRvD..75d3514M}%
  \BibitemOpen
  \bibfield  {author} {\bibinfo {author} {\bibfnamefont {P.}~\bibnamefont
  {{McDonald}}},\ }\bibfield  {title} {\enquote {\bibinfo {title} {{Dark matter
  clustering: A simple renormalization group approach}},}\ }\href {\doibase
  10.1103/PhysRevD.75.043514} {\bibfield  {journal} {\bibinfo  {journal}
  {\prd}\ }\textbf {\bibinfo {volume} {75}},\ \bibinfo {pages} {043514--+}
  (\bibinfo {year} {2007})}\BibitemShut {NoStop}%
\bibitem [{\citenamefont {{McDonald}}(2011)}]{2011JCAP...04..032M}%
  \BibitemOpen
  \bibfield  {author} {\bibinfo {author} {\bibfnamefont {P.}~\bibnamefont
  {{McDonald}}},\ }\bibfield  {title} {\enquote {\bibinfo {title} {{How to
  generate a significant effective temperature for cold dark matter, from first
  principles}},}\ }\href {\doibase 10.1088/1475-7516/2011/04/032} {\bibfield
  {journal} {\bibinfo  {journal} {\jcap}\ }\textbf {\bibinfo {volume} {4}},\
  \bibinfo {pages} {32--+} (\bibinfo {year} {2011})},\ \Eprint
  {http://arxiv.org/abs/0910.1002} {arXiv:0910.1002 [astro-ph.CO]} \BibitemShut
  {NoStop}%
\bibitem [{\citenamefont {{Seljak}}\ and\ \citenamefont
  {{McDonald}}(2011)}]{2011JCAP...11..039S}%
  \BibitemOpen
  \bibfield  {author} {\bibinfo {author} {\bibfnamefont {U.}~\bibnamefont
  {{Seljak}}}\ and\ \bibinfo {author} {\bibfnamefont {P.}~\bibnamefont
  {{McDonald}}},\ }\bibfield  {title} {\enquote {\bibinfo {title}
  {{Distribution function approach to redshift space distortions}},}\ }\href
  {\doibase 10.1088/1475-7516/2011/11/039} {\bibfield  {journal} {\bibinfo
  {journal} {\jcap}\ }\textbf {\bibinfo {volume} {11}},\ \bibinfo {eid} {039}
  (\bibinfo {year} {2011})},\ \Eprint {http://arxiv.org/abs/1109.1888}
  {arXiv:1109.1888 [astro-ph.CO]} \BibitemShut {NoStop}%
\bibitem [{\citenamefont {{Baumann}}\ \emph {et~al.}(2012)\citenamefont
  {{Baumann}}, \citenamefont {{Nicolis}}, \citenamefont {{Senatore}},\ and\
  \citenamefont {{Zaldarriaga}}}]{2012JCAP...07..051B}%
  \BibitemOpen
  \bibfield  {author} {\bibinfo {author} {\bibfnamefont {D.}~\bibnamefont
  {{Baumann}}}, \bibinfo {author} {\bibfnamefont {A.}~\bibnamefont
  {{Nicolis}}}, \bibinfo {author} {\bibfnamefont {L.}~\bibnamefont
  {{Senatore}}}, \ and\ \bibinfo {author} {\bibfnamefont {M.}~\bibnamefont
  {{Zaldarriaga}}},\ }\bibfield  {title} {\enquote {\bibinfo {title}
  {{Cosmological non-linearities as an effective fluid}},}\ }\href {\doibase
  10.1088/1475-7516/2012/07/051} {\bibfield  {journal} {\bibinfo  {journal}
  {\jcap}\ }\textbf {\bibinfo {volume} {7}},\ \bibinfo {eid} {051} (\bibinfo
  {year} {2012})},\ \Eprint {http://arxiv.org/abs/1004.2488} {arXiv:1004.2488
  [astro-ph.CO]} \BibitemShut {NoStop}%
\bibitem [{\citenamefont {{Carrasco}}\ \emph {et~al.}(2012)\citenamefont
  {{Carrasco}}, \citenamefont {{Hertzberg}},\ and\ \citenamefont
  {{Senatore}}}]{2012JHEP...09..082C}%
  \BibitemOpen
  \bibfield  {author} {\bibinfo {author} {\bibfnamefont {J.~J.~M.}\
  \bibnamefont {{Carrasco}}}, \bibinfo {author} {\bibfnamefont {M.~P.}\
  \bibnamefont {{Hertzberg}}}, \ and\ \bibinfo {author} {\bibfnamefont
  {L.}~\bibnamefont {{Senatore}}},\ }\bibfield  {title} {\enquote {\bibinfo
  {title} {{The effective field theory of cosmological large scale
  structures}},}\ }\href {\doibase 10.1007/JHEP09(2012)082} {\bibfield
  {journal} {\bibinfo  {journal} {Journal of High Energy Physics}\ }\textbf
  {\bibinfo {volume} {9}},\ \bibinfo {pages} {82} (\bibinfo {year} {2012})},\
  \Eprint {http://arxiv.org/abs/1206.2926} {arXiv:1206.2926 [astro-ph.CO]}
  \BibitemShut {NoStop}%
\bibitem [{\citenamefont {{Hertzberg}}(2014)}]{2014PhRvD..89d3521H}%
  \BibitemOpen
  \bibfield  {author} {\bibinfo {author} {\bibfnamefont {M.~P.}\ \bibnamefont
  {{Hertzberg}}},\ }\bibfield  {title} {\enquote {\bibinfo {title} {{Effective
  field theory of dark matter and structure formation: Semianalytical
  results}},}\ }\href {\doibase 10.1103/PhysRevD.89.043521} {\bibfield
  {journal} {\bibinfo  {journal} {\prd}\ }\textbf {\bibinfo {volume} {89}},\
  \bibinfo {eid} {043521} (\bibinfo {year} {2014})}\BibitemShut {NoStop}%
\bibitem [{\citenamefont {{McDonald}}(2014)}]{2014arXiv1403.7235M}%
  \BibitemOpen
  \bibfield  {author} {\bibinfo {author} {\bibfnamefont {P.}~\bibnamefont
  {{McDonald}}},\ }\bibfield  {title} {\enquote {\bibinfo {title} {{What the
  ''simple renormalization group'' approach to dark matter clustering really
  was}},}\ }\href@noop {} {\bibfield  {journal} {\bibinfo  {journal} {ArXiv
  e-prints}\ } (\bibinfo {year} {2014})},\ \Eprint
  {http://arxiv.org/abs/1403.7235} {arXiv:1403.7235} \BibitemShut {NoStop}%
\bibitem [{\citenamefont {{Blas}}\ \emph {et~al.}(2015)\citenamefont {{Blas}},
  \citenamefont {{Floerchinger}}, \citenamefont {{Garny}}, \citenamefont
  {{Tetradis}},\ and\ \citenamefont {{Wiedemann}}}]{2015JCAP...11..049B}%
  \BibitemOpen
  \bibfield  {author} {\bibinfo {author} {\bibfnamefont {D.}~\bibnamefont
  {{Blas}}}, \bibinfo {author} {\bibfnamefont {S.}~\bibnamefont
  {{Floerchinger}}}, \bibinfo {author} {\bibfnamefont {M.}~\bibnamefont
  {{Garny}}}, \bibinfo {author} {\bibfnamefont {N.}~\bibnamefont {{Tetradis}}},
  \ and\ \bibinfo {author} {\bibfnamefont {U.~A.}\ \bibnamefont
  {{Wiedemann}}},\ }\bibfield  {title} {\enquote {\bibinfo {title} {{Large
  scale structure from viscous dark matter}},}\ }\href {\doibase
  10.1088/1475-7516/2015/11/049} {\bibfield  {journal} {\bibinfo  {journal}
  {\jcap}\ }\textbf {\bibinfo {volume} {11}},\ \bibinfo {eid} {049} (\bibinfo
  {year} {2015})},\ \Eprint {http://arxiv.org/abs/1507.06665}
  {arXiv:1507.06665} \BibitemShut {NoStop}%
\bibitem [{\citenamefont {{Rigopoulos}}(2015)}]{2015JCAP...01..014R}%
  \BibitemOpen
  \bibfield  {author} {\bibinfo {author} {\bibfnamefont {G.}~\bibnamefont
  {{Rigopoulos}}},\ }\bibfield  {title} {\enquote {\bibinfo {title} {{The
  adhesion model as a field theory for cosmological clustering}},}\ }\href
  {\doibase 10.1088/1475-7516/2015/01/014} {\bibfield  {journal} {\bibinfo
  {journal} {\jcap}\ }\textbf {\bibinfo {volume} {1}},\ \bibinfo {eid} {014}
  (\bibinfo {year} {2015})},\ \Eprint {http://arxiv.org/abs/1404.7283}
  {arXiv:1404.7283} \BibitemShut {NoStop}%
\bibitem [{\citenamefont {{F{\"u}hrer}}\ and\ \citenamefont
  {{Rigopoulos}}(2016)}]{2016JCAP...02..032F}%
  \BibitemOpen
  \bibfield  {author} {\bibinfo {author} {\bibfnamefont {F.}~\bibnamefont
  {{F{\"u}hrer}}}\ and\ \bibinfo {author} {\bibfnamefont {G.}~\bibnamefont
  {{Rigopoulos}}},\ }\bibfield  {title} {\enquote {\bibinfo {title}
  {{Renormalizing a viscous fluid model for large scale structure
  formation}},}\ }\href {\doibase 10.1088/1475-7516/2016/02/032} {\bibfield
  {journal} {\bibinfo  {journal} {\jcap}\ }\textbf {\bibinfo {volume} {2}},\
  \bibinfo {eid} {032} (\bibinfo {year} {2016})},\ \Eprint
  {http://arxiv.org/abs/1509.03073} {arXiv:1509.03073} \BibitemShut {NoStop}%
\bibitem [{\citenamefont {{Bartelmann}}\ \emph {et~al.}(2016)\citenamefont
  {{Bartelmann}}, \citenamefont {{Fabis}}, \citenamefont {{Berg}},
  \citenamefont {{Kozlikin}}, \citenamefont {{Lilow}},\ and\ \citenamefont
  {{Viermann}}}]{2016NJPh...18d3020B}%
  \BibitemOpen
  \bibfield  {author} {\bibinfo {author} {\bibfnamefont {M.}~\bibnamefont
  {{Bartelmann}}}, \bibinfo {author} {\bibfnamefont {F.}~\bibnamefont
  {{Fabis}}}, \bibinfo {author} {\bibfnamefont {D.}~\bibnamefont {{Berg}}},
  \bibinfo {author} {\bibfnamefont {E.}~\bibnamefont {{Kozlikin}}}, \bibinfo
  {author} {\bibfnamefont {R.}~\bibnamefont {{Lilow}}}, \ and\ \bibinfo
  {author} {\bibfnamefont {C.}~\bibnamefont {{Viermann}}},\ }\bibfield  {title}
  {\enquote {\bibinfo {title} {{A microscopic, non-equilibrium, statistical
  field theory for cosmic structure formation}},}\ }\href {\doibase
  10.1088/1367-2630/18/4/043020} {\bibfield  {journal} {\bibinfo  {journal}
  {New Journal of Physics}\ }\textbf {\bibinfo {volume} {18}},\ \bibinfo {eid}
  {043020} (\bibinfo {year} {2016})}\BibitemShut {NoStop}%
\bibitem [{\citenamefont {{Schmittfull}}\ \emph {et~al.}(2016)\citenamefont
  {{Schmittfull}}, \citenamefont {{Vlah}},\ and\ \citenamefont
  {{McDonald}}}]{2016PhRvD..93j3528S}%
  \BibitemOpen
  \bibfield  {author} {\bibinfo {author} {\bibfnamefont {M.}~\bibnamefont
  {{Schmittfull}}}, \bibinfo {author} {\bibfnamefont {Z.}~\bibnamefont
  {{Vlah}}}, \ and\ \bibinfo {author} {\bibfnamefont {P.}~\bibnamefont
  {{McDonald}}},\ }\bibfield  {title} {\enquote {\bibinfo {title} {{Fast large
  scale structure perturbation theory using one-dimensional fast Fourier
  transforms}},}\ }\href {\doibase 10.1103/PhysRevD.93.103528} {\bibfield
  {journal} {\bibinfo  {journal} {\prd}\ }\textbf {\bibinfo {volume} {93}},\
  \bibinfo {eid} {103528} (\bibinfo {year} {2016})},\ \Eprint
  {http://arxiv.org/abs/1603.04405} {arXiv:1603.04405} \BibitemShut {NoStop}%
\bibitem [{\citenamefont {{Feng}}\ \emph {et~al.}(2016)\citenamefont {{Feng}},
  \citenamefont {{Chu}}, \citenamefont {{Seljak}},\ and\ \citenamefont
  {{McDonald}}}]{2016MNRAS.463.2273F}%
  \BibitemOpen
  \bibfield  {author} {\bibinfo {author} {\bibfnamefont {Y.}~\bibnamefont
  {{Feng}}}, \bibinfo {author} {\bibfnamefont {M.-Y.}\ \bibnamefont {{Chu}}},
  \bibinfo {author} {\bibfnamefont {U.}~\bibnamefont {{Seljak}}}, \ and\
  \bibinfo {author} {\bibfnamefont {P.}~\bibnamefont {{McDonald}}},\ }\bibfield
   {title} {\enquote {\bibinfo {title} {{FASTPM: a new scheme for fast
  simulations of dark matter and haloes}},}\ }\href {\doibase
  10.1093/mnras/stw2123} {\bibfield  {journal} {\bibinfo  {journal} {\mnras}\
  }\textbf {\bibinfo {volume} {463}},\ \bibinfo {pages} {2273--2286} (\bibinfo
  {year} {2016})},\ \Eprint {http://arxiv.org/abs/1603.00476}
  {arXiv:1603.00476} \BibitemShut {NoStop}%
\bibitem [{\citenamefont {{McDonald}}\ and\ \citenamefont
  {{Vlah}}(2018)}]{2018PhRvD..97b3508M}%
  \BibitemOpen
  \bibfield  {author} {\bibinfo {author} {\bibfnamefont {Patrick}\ \bibnamefont
  {{McDonald}}}\ and\ \bibinfo {author} {\bibfnamefont {Zvonimir}\ \bibnamefont
  {{Vlah}}},\ }\bibfield  {title} {\enquote {\bibinfo {title} {{Large-scale
  structure perturbation theory without losing stream crossing}},}\ }\href
  {\doibase 10.1103/PhysRevD.97.023508} {\bibfield  {journal} {\bibinfo
  {journal} {\prd}\ }\textbf {\bibinfo {volume} {97}},\ \bibinfo {eid} {023508}
  (\bibinfo {year} {2018})}\BibitemShut {NoStop}%
\bibitem [{\citenamefont {{Jelic-Cizmek}}\ \emph {et~al.}(2018)\citenamefont
  {{Jelic-Cizmek}}, \citenamefont {{Lepori}}, \citenamefont {{Adamek}},\ and\
  \citenamefont {{Durrer}}}]{2018JCAP...09..006J}%
  \BibitemOpen
  \bibfield  {author} {\bibinfo {author} {\bibfnamefont {Goran}\ \bibnamefont
  {{Jelic-Cizmek}}}, \bibinfo {author} {\bibfnamefont {Francesca}\ \bibnamefont
  {{Lepori}}}, \bibinfo {author} {\bibfnamefont {Julian}\ \bibnamefont
  {{Adamek}}}, \ and\ \bibinfo {author} {\bibfnamefont {Ruth}\ \bibnamefont
  {{Durrer}}},\ }\bibfield  {title} {\enquote {\bibinfo {title} {{The
  generation of vorticity in cosmological N-body simulations}},}\ }\href
  {\doibase 10.1088/1475-7516/2018/09/006} {\bibfield  {journal} {\bibinfo
  {journal} {Journal of Cosmology and Astro-Particle Physics}\ }\textbf
  {\bibinfo {volume} {2018}},\ \bibinfo {eid} {006} (\bibinfo {year}
  {2018})}\BibitemShut {NoStop}%
\bibitem [{\citenamefont {{Pietroni}}(2018)}]{2018JCAP...06..028P}%
  \BibitemOpen
  \bibfield  {author} {\bibinfo {author} {\bibfnamefont {Massimo}\ \bibnamefont
  {{Pietroni}}},\ }\bibfield  {title} {\enquote {\bibinfo {title} {{Structure
  formation beyond shell-crossing: nonperturbative expansions and late-time
  attractors}},}\ }\href {\doibase 10.1088/1475-7516/2018/06/028} {\bibfield
  {journal} {\bibinfo  {journal} {Journal of Cosmology and Astro-Particle
  Physics}\ }\textbf {\bibinfo {volume} {2018}},\ \bibinfo {eid} {028}
  (\bibinfo {year} {2018})}\BibitemShut {NoStop}%
\bibitem [{\citenamefont {{Pajer}}\ and\ \citenamefont {{van der
  Woude}}(2018)}]{2018JCAP...05..039P}%
  \BibitemOpen
  \bibfield  {author} {\bibinfo {author} {\bibfnamefont {Enrico}\ \bibnamefont
  {{Pajer}}}\ and\ \bibinfo {author} {\bibfnamefont {Drian}\ \bibnamefont {{van
  der Woude}}},\ }\bibfield  {title} {\enquote {\bibinfo {title} {{Divergence
  of perturbation theory in large scale structures}},}\ }\href {\doibase
  10.1088/1475-7516/2018/05/039} {\bibfield  {journal} {\bibinfo  {journal}
  {Journal of Cosmology and Astro-Particle Physics}\ }\textbf {\bibinfo
  {volume} {2018}},\ \bibinfo {eid} {039} (\bibinfo {year} {2018})}\BibitemShut
  {NoStop}%
\bibitem [{\citenamefont {{Fabis}}\ \emph {et~al.}(2018)\citenamefont
  {{Fabis}}, \citenamefont {{Kozlikin}}, \citenamefont {{Lilow}},\ and\
  \citenamefont {{Bartelmann}}}]{2018JSMTE..04.3214F}%
  \BibitemOpen
  \bibfield  {author} {\bibinfo {author} {\bibfnamefont {Felix}\ \bibnamefont
  {{Fabis}}}, \bibinfo {author} {\bibfnamefont {Elena}\ \bibnamefont
  {{Kozlikin}}}, \bibinfo {author} {\bibfnamefont {Robert}\ \bibnamefont
  {{Lilow}}}, \ and\ \bibinfo {author} {\bibfnamefont {Matthias}\ \bibnamefont
  {{Bartelmann}}},\ }\bibfield  {title} {\enquote {\bibinfo {title} {{Kinetic
  field theory: exact free evolution of Gaussian phase-space correlations}},}\
  }\href {\doibase 10.1088/1742-5468/aab850} {\bibfield  {journal} {\bibinfo
  {journal} {Journal of Statistical Mechanics: Theory and Experiment}\ }\textbf
  {\bibinfo {volume} {4}},\ \bibinfo {pages} {043214} (\bibinfo {year}
  {2018})}\BibitemShut {NoStop}%
\bibitem [{\citenamefont {{Kopp}}\ \emph {et~al.}(2017)\citenamefont {{Kopp}},
  \citenamefont {{Vattis}},\ and\ \citenamefont
  {{Skordis}}}]{2017PhRvD..96l3532K}%
  \BibitemOpen
  \bibfield  {author} {\bibinfo {author} {\bibfnamefont {Michael}\ \bibnamefont
  {{Kopp}}}, \bibinfo {author} {\bibfnamefont {Kyriakos}\ \bibnamefont
  {{Vattis}}}, \ and\ \bibinfo {author} {\bibfnamefont {Constantinos}\
  \bibnamefont {{Skordis}}},\ }\bibfield  {title} {\enquote {\bibinfo {title}
  {{Solving the Vlasov equation in two spatial dimensions with the
  Schr{\"o}dinger method}},}\ }\href {\doibase 10.1103/PhysRevD.96.123532}
  {\bibfield  {journal} {\bibinfo  {journal} {\prd}\ }\textbf {\bibinfo
  {volume} {96}},\ \bibinfo {eid} {123532} (\bibinfo {year}
  {2017})}\BibitemShut {NoStop}%
\bibitem [{\citenamefont {{Taruya}}\ \emph {et~al.}(2018)\citenamefont
  {{Taruya}}, \citenamefont {{Nishimichi}},\ and\ \citenamefont
  {{Jeong}}}]{2018arXiv180704215T}%
  \BibitemOpen
  \bibfield  {author} {\bibinfo {author} {\bibfnamefont {Atsushi}\ \bibnamefont
  {{Taruya}}}, \bibinfo {author} {\bibfnamefont {Takahiro}\ \bibnamefont
  {{Nishimichi}}}, \ and\ \bibinfo {author} {\bibfnamefont {Donghui}\
  \bibnamefont {{Jeong}}},\ }\bibfield  {title} {\enquote {\bibinfo {title}
  {{GridSPT: Grid-based calculation for perturbation theory of large-scale
  structure}},}\ }\href@noop {} {\bibfield  {journal} {\bibinfo  {journal}
  {ArXiv e-prints}\ ,\ \bibinfo {eid} {arXiv:1807.04215}} (\bibinfo {year}
  {2018})},\ \Eprint {http://arxiv.org/abs/1807.04215} {arXiv:1807.04215
  [astro-ph.CO]} \BibitemShut {NoStop}%
\bibitem [{\citenamefont {{Jenkins}}\ \emph {et~al.}(2014)\citenamefont
  {{Jenkins}}, \citenamefont {{Manohar}}, \citenamefont {{Waalewijn}},\ and\
  \citenamefont {{Yadav}}}]{2014JCAP...09..024J}%
  \BibitemOpen
  \bibfield  {author} {\bibinfo {author} {\bibfnamefont {E.~E.}\ \bibnamefont
  {{Jenkins}}}, \bibinfo {author} {\bibfnamefont {A.~V.}\ \bibnamefont
  {{Manohar}}}, \bibinfo {author} {\bibfnamefont {W.~J.}\ \bibnamefont
  {{Waalewijn}}}, \ and\ \bibinfo {author} {\bibfnamefont {A.~P.~S.}\
  \bibnamefont {{Yadav}}},\ }\bibfield  {title} {\enquote {\bibinfo {title}
  {{Higher-order gravitational lensing reconstruction using Feynman
  diagrams}},}\ }\href {\doibase 10.1088/1475-7516/2014/09/024} {\bibfield
  {journal} {\bibinfo  {journal} {\jcap}\ }\textbf {\bibinfo {volume} {9}},\
  \bibinfo {eid} {024} (\bibinfo {year} {2014})},\ \Eprint
  {http://arxiv.org/abs/1403.4607} {arXiv:1403.4607} \BibitemShut {NoStop}%
\bibitem [{\citenamefont {{Madhavacheril}}\ \emph {et~al.}(2015)\citenamefont
  {{Madhavacheril}}, \citenamefont {{McDonald}}, \citenamefont {{Sehgal}},\
  and\ \citenamefont {{Slosar}}}]{2015JCAP...01..022M}%
  \BibitemOpen
  \bibfield  {author} {\bibinfo {author} {\bibfnamefont {M.~S.}\ \bibnamefont
  {{Madhavacheril}}}, \bibinfo {author} {\bibfnamefont {P.}~\bibnamefont
  {{McDonald}}}, \bibinfo {author} {\bibfnamefont {N.}~\bibnamefont
  {{Sehgal}}}, \ and\ \bibinfo {author} {\bibfnamefont {A.}~\bibnamefont
  {{Slosar}}},\ }\bibfield  {title} {\enquote {\bibinfo {title} {{Building
  unbiased estimators from non-Gaussian likelihoods with application to shear
  estimation}},}\ }\href {\doibase 10.1088/1475-7516/2015/01/022} {\bibfield
  {journal} {\bibinfo  {journal} {\jcap}\ }\textbf {\bibinfo {volume} {1}},\
  \bibinfo {eid} {022} (\bibinfo {year} {2015})},\ \Eprint
  {http://arxiv.org/abs/1407.1906} {arXiv:1407.1906} \BibitemShut {NoStop}%
\bibitem [{\citenamefont {{Di Dio}}(2017)}]{2017JCAP...03..016D}%
  \BibitemOpen
  \bibfield  {author} {\bibinfo {author} {\bibfnamefont {E.}~\bibnamefont {{Di
  Dio}}},\ }\bibfield  {title} {\enquote {\bibinfo {title} {{Lensing smoothing
  of BAO wiggles}},}\ }\href {\doibase 10.1088/1475-7516/2017/03/016}
  {\bibfield  {journal} {\bibinfo  {journal} {\jcap}\ }\textbf {\bibinfo
  {volume} {3}},\ \bibinfo {eid} {016} (\bibinfo {year} {2017})},\ \Eprint
  {http://arxiv.org/abs/1609.09044} {arXiv:1609.09044} \BibitemShut {NoStop}%
\bibitem [{\citenamefont {{McDonald}}(2006)}]{2006PhRvD..74j3512M}%
  \BibitemOpen
  \bibfield  {author} {\bibinfo {author} {\bibfnamefont {P.}~\bibnamefont
  {{McDonald}}},\ }\bibfield  {title} {\enquote {\bibinfo {title} {{Clustering
  of dark matter tracers: Renormalizing the bias parameters}},}\ }\href
  {\doibase 10.1103/PhysRevD.74.103512} {\bibfield  {journal} {\bibinfo
  {journal} {\prd}\ }\textbf {\bibinfo {volume} {74}},\ \bibinfo {pages}
  {103512--+} (\bibinfo {year} {2006})}\BibitemShut {NoStop}%
\bibitem [{\citenamefont {{Smith}}\ \emph
  {et~al.}(2007{\natexlab{b}})\citenamefont {{Smith}}, \citenamefont
  {{Scoccimarro}},\ and\ \citenamefont {{Sheth}}}]{2007PhRvD..75f3512S}%
  \BibitemOpen
  \bibfield  {author} {\bibinfo {author} {\bibfnamefont {R.~E.}\ \bibnamefont
  {{Smith}}}, \bibinfo {author} {\bibfnamefont {R.}~\bibnamefont
  {{Scoccimarro}}}, \ and\ \bibinfo {author} {\bibfnamefont {R.~K.}\
  \bibnamefont {{Sheth}}},\ }\bibfield  {title} {\enquote {\bibinfo {title}
  {{Scale dependence of halo and galaxy bias: Effects in real space}},}\ }\href
  {\doibase 10.1103/PhysRevD.75.063512} {\bibfield  {journal} {\bibinfo
  {journal} {\prd}\ }\textbf {\bibinfo {volume} {75}},\ \bibinfo {pages}
  {063512--+} (\bibinfo {year} {2007}{\natexlab{b}})},\ \Eprint
  {http://arxiv.org/abs/arXiv:astro-ph/0609547} {arXiv:astro-ph/0609547}
  \BibitemShut {NoStop}%
\bibitem [{\citenamefont {{Matsubara}}(2008)}]{2008PhRvD..78h3519M}%
  \BibitemOpen
  \bibfield  {author} {\bibinfo {author} {\bibfnamefont {T.}~\bibnamefont
  {{Matsubara}}},\ }\bibfield  {title} {\enquote {\bibinfo {title} {{Nonlinear
  perturbation theory with halo bias and redshift-space distortions via the
  Lagrangian picture}},}\ }\href {\doibase 10.1103/PhysRevD.78.083519}
  {\bibfield  {journal} {\bibinfo  {journal} {\prd}\ }\textbf {\bibinfo
  {volume} {78}},\ \bibinfo {pages} {083519--+} (\bibinfo {year}
  {2008})}\BibitemShut {NoStop}%
\bibitem [{\citenamefont {{McDonald}}\ and\ \citenamefont
  {{Roy}}(2009)}]{2009JCAP...08..020M}%
  \BibitemOpen
  \bibfield  {author} {\bibinfo {author} {\bibfnamefont {P.}~\bibnamefont
  {{McDonald}}}\ and\ \bibinfo {author} {\bibfnamefont {A.}~\bibnamefont
  {{Roy}}},\ }\bibfield  {title} {\enquote {\bibinfo {title} {{Clustering of
  dark matter tracers: generalizing bias for the coming era of precision
  LSS}},}\ }\href {\doibase 10.1088/1475-7516/2009/08/020} {\bibfield
  {journal} {\bibinfo  {journal} {\jcap}\ }\textbf {\bibinfo {volume} {8}},\
  \bibinfo {eid} {020} (\bibinfo {year} {2009})},\ \Eprint
  {http://arxiv.org/abs/0902.0991} {arXiv:0902.0991 [astro-ph.CO]} \BibitemShut
  {NoStop}%
\bibitem [{\citenamefont {{Jeong}}\ and\ \citenamefont
  {{Komatsu}}(2009)}]{2009ApJ...691..569J}%
  \BibitemOpen
  \bibfield  {author} {\bibinfo {author} {\bibfnamefont {D.}~\bibnamefont
  {{Jeong}}}\ and\ \bibinfo {author} {\bibfnamefont {E.}~\bibnamefont
  {{Komatsu}}},\ }\bibfield  {title} {\enquote {\bibinfo {title} {{Perturbation
  Theory Reloaded. II. Nonlinear Bias, Baryon Acoustic Oscillations, and
  Millennium Simulation in Real Space}},}\ }\href {\doibase
  10.1088/0004-637X/691/1/569} {\bibfield  {journal} {\bibinfo  {journal}
  {\apj}\ }\textbf {\bibinfo {volume} {691}},\ \bibinfo {pages} {569--595}
  (\bibinfo {year} {2009})},\ \Eprint {http://arxiv.org/abs/0805.2632}
  {arXiv:0805.2632} \BibitemShut {NoStop}%
\bibitem [{\citenamefont {{Matsubara}}(2011)}]{2011PhRvD..83h3518M}%
  \BibitemOpen
  \bibfield  {author} {\bibinfo {author} {\bibfnamefont {T.}~\bibnamefont
  {{Matsubara}}},\ }\bibfield  {title} {\enquote {\bibinfo {title} {{Nonlinear
  perturbation theory integrated with nonlocal bias, redshift-space
  distortions, and primordial non-Gaussianity}},}\ }\href {\doibase
  10.1103/PhysRevD.83.083518} {\bibfield  {journal} {\bibinfo  {journal}
  {\prd}\ }\textbf {\bibinfo {volume} {83}},\ \bibinfo {eid} {083518} (\bibinfo
  {year} {2011})},\ \Eprint {http://arxiv.org/abs/1102.4619} {arXiv:1102.4619
  [astro-ph.CO]} \BibitemShut {NoStop}%
\bibitem [{\citenamefont {{Elia}}\ \emph {et~al.}(2011)\citenamefont {{Elia}},
  \citenamefont {{Kulkarni}}, \citenamefont {{Porciani}}, \citenamefont
  {{Pietroni}},\ and\ \citenamefont {{Matarrese}}}]{2011MNRAS.416.1703E}%
  \BibitemOpen
  \bibfield  {author} {\bibinfo {author} {\bibfnamefont {A.}~\bibnamefont
  {{Elia}}}, \bibinfo {author} {\bibfnamefont {S.}~\bibnamefont {{Kulkarni}}},
  \bibinfo {author} {\bibfnamefont {C.}~\bibnamefont {{Porciani}}}, \bibinfo
  {author} {\bibfnamefont {M.}~\bibnamefont {{Pietroni}}}, \ and\ \bibinfo
  {author} {\bibfnamefont {S.}~\bibnamefont {{Matarrese}}},\ }\bibfield
  {title} {\enquote {\bibinfo {title} {{Modelling the clustering of dark matter
  haloes in resummed perturbation theories}},}\ }\href {\doibase
  10.1111/j.1365-2966.2011.18761.x} {\bibfield  {journal} {\bibinfo  {journal}
  {\mnras}\ }\textbf {\bibinfo {volume} {416}},\ \bibinfo {pages} {1703--1716}
  (\bibinfo {year} {2011})},\ \Eprint {http://arxiv.org/abs/1012.4833}
  {arXiv:1012.4833 [astro-ph.CO]} \BibitemShut {NoStop}%
\bibitem [{\citenamefont {{Chan}}\ and\ \citenamefont
  {{Scoccimarro}}(2012)}]{2012PhRvD..86j3519C}%
  \BibitemOpen
  \bibfield  {author} {\bibinfo {author} {\bibfnamefont {K.~C.}\ \bibnamefont
  {{Chan}}}\ and\ \bibinfo {author} {\bibfnamefont {R.}~\bibnamefont
  {{Scoccimarro}}},\ }\bibfield  {title} {\enquote {\bibinfo {title} {{Halo
  sampling, local bias, and loop corrections}},}\ }\href {\doibase
  10.1103/PhysRevD.86.103519} {\bibfield  {journal} {\bibinfo  {journal}
  {\prd}\ }\textbf {\bibinfo {volume} {86}},\ \bibinfo {eid} {103519} (\bibinfo
  {year} {2012})},\ \Eprint {http://arxiv.org/abs/1204.5770} {arXiv:1204.5770
  [astro-ph.CO]} \BibitemShut {NoStop}%
\bibitem [{\citenamefont {{Baldauf}}\ \emph {et~al.}(2012)\citenamefont
  {{Baldauf}}, \citenamefont {{Seljak}}, \citenamefont {{Desjacques}},\ and\
  \citenamefont {{McDonald}}}]{2012PhRvD..86h3540B}%
  \BibitemOpen
  \bibfield  {author} {\bibinfo {author} {\bibfnamefont {T.}~\bibnamefont
  {{Baldauf}}}, \bibinfo {author} {\bibfnamefont {U.}~\bibnamefont {{Seljak}}},
  \bibinfo {author} {\bibfnamefont {V.}~\bibnamefont {{Desjacques}}}, \ and\
  \bibinfo {author} {\bibfnamefont {P.}~\bibnamefont {{McDonald}}},\ }\bibfield
   {title} {\enquote {\bibinfo {title} {{Evidence for quadratic tidal tensor
  bias from the halo bispectrum}},}\ }\href {\doibase
  10.1103/PhysRevD.86.083540} {\bibfield  {journal} {\bibinfo  {journal}
  {\prd}\ }\textbf {\bibinfo {volume} {86}},\ \bibinfo {eid} {083540} (\bibinfo
  {year} {2012})},\ \Eprint {http://arxiv.org/abs/1201.4827} {arXiv:1201.4827
  [astro-ph.CO]} \BibitemShut {NoStop}%
\bibitem [{\citenamefont {{Nishimichi}}\ and\ \citenamefont
  {{Oka}}(2014)}]{2014MNRAS.444.1400N}%
  \BibitemOpen
  \bibfield  {author} {\bibinfo {author} {\bibfnamefont {T.}~\bibnamefont
  {{Nishimichi}}}\ and\ \bibinfo {author} {\bibfnamefont {A.}~\bibnamefont
  {{Oka}}},\ }\bibfield  {title} {\enquote {\bibinfo {title} {{Simulating the
  anisotropic clustering of luminous red galaxies with subhaloes: a direct
  confrontation with observation and cosmological implications}},}\ }\href
  {\doibase 10.1093/mnras/stu1528} {\bibfield  {journal} {\bibinfo  {journal}
  {\mnras}\ }\textbf {\bibinfo {volume} {444}},\ \bibinfo {pages} {1400--1418}
  (\bibinfo {year} {2014})},\ \Eprint {http://arxiv.org/abs/1310.2672}
  {arXiv:1310.2672} \BibitemShut {NoStop}%
\bibitem [{\citenamefont {{Saito}}\ \emph {et~al.}(2014)\citenamefont
  {{Saito}}, \citenamefont {{Baldauf}}, \citenamefont {{Vlah}}, \citenamefont
  {{Seljak}}, \citenamefont {{Okumura}},\ and\ \citenamefont
  {{McDonald}}}]{2014PhRvD..90l3522S}%
  \BibitemOpen
  \bibfield  {author} {\bibinfo {author} {\bibfnamefont {S.}~\bibnamefont
  {{Saito}}}, \bibinfo {author} {\bibfnamefont {T.}~\bibnamefont {{Baldauf}}},
  \bibinfo {author} {\bibfnamefont {Z.}~\bibnamefont {{Vlah}}}, \bibinfo
  {author} {\bibfnamefont {U.}~\bibnamefont {{Seljak}}}, \bibinfo {author}
  {\bibfnamefont {T.}~\bibnamefont {{Okumura}}}, \ and\ \bibinfo {author}
  {\bibfnamefont {P.}~\bibnamefont {{McDonald}}},\ }\bibfield  {title}
  {\enquote {\bibinfo {title} {{Understanding higher-order nonlocal halo bias
  at large scales by combining the power spectrum with the bispectrum}},}\
  }\href {\doibase 10.1103/PhysRevD.90.123522} {\bibfield  {journal} {\bibinfo
  {journal} {\prd}\ }\textbf {\bibinfo {volume} {90}},\ \bibinfo {eid} {123522}
  (\bibinfo {year} {2014})},\ \Eprint {http://arxiv.org/abs/1405.1447}
  {arXiv:1405.1447} \BibitemShut {NoStop}%
\bibitem [{\citenamefont {{Desjacques}}\ \emph {et~al.}(2016)\citenamefont
  {{Desjacques}}, \citenamefont {{Jeong}},\ and\ \citenamefont
  {{Schmidt}}}]{2016arXiv161109787D}%
  \BibitemOpen
  \bibfield  {author} {\bibinfo {author} {\bibfnamefont {V.}~\bibnamefont
  {{Desjacques}}}, \bibinfo {author} {\bibfnamefont {D.}~\bibnamefont
  {{Jeong}}}, \ and\ \bibinfo {author} {\bibfnamefont {F.}~\bibnamefont
  {{Schmidt}}},\ }\bibfield  {title} {\enquote {\bibinfo {title} {{Large-Scale
  Galaxy Bias}},}\ }\href@noop {} {\bibfield  {journal} {\bibinfo  {journal}
  {ArXiv e-prints}\ } (\bibinfo {year} {2016})},\ \Eprint
  {http://arxiv.org/abs/1611.09787} {arXiv:1611.09787} \BibitemShut {NoStop}%
\bibitem [{\citenamefont {{Hand}}\ \emph {et~al.}(2017)\citenamefont {{Hand}},
  \citenamefont {{Seljak}}, \citenamefont {{Beutler}},\ and\ \citenamefont
  {{Vlah}}}]{2017arXiv170602362H}%
  \BibitemOpen
  \bibfield  {author} {\bibinfo {author} {\bibfnamefont {N.}~\bibnamefont
  {{Hand}}}, \bibinfo {author} {\bibfnamefont {U.}~\bibnamefont {{Seljak}}},
  \bibinfo {author} {\bibfnamefont {F.}~\bibnamefont {{Beutler}}}, \ and\
  \bibinfo {author} {\bibfnamefont {Z.}~\bibnamefont {{Vlah}}},\ }\bibfield
  {title} {\enquote {\bibinfo {title} {{Extending the modeling of the
  anisotropic galaxy power spectrum to $k = 0.4 \ h\mathrm{Mpc}^{-1}$}},}\
  }\href@noop {} {\bibfield  {journal} {\bibinfo  {journal} {ArXiv e-prints}\ }
  (\bibinfo {year} {2017})},\ \Eprint {http://arxiv.org/abs/1706.02362}
  {arXiv:1706.02362} \BibitemShut {NoStop}%
\bibitem [{\citenamefont {{Chuang}}\ \emph {et~al.}(2017)\citenamefont
  {{Chuang}}, \citenamefont {{Kitaura}}, \citenamefont {{Liang}}, \citenamefont
  {{Font-Ribera}}, \citenamefont {{Zhao}}, \citenamefont {{McDonald}},\ and\
  \citenamefont {{Tao}}}]{2017PhRvD..95f3528C}%
  \BibitemOpen
  \bibfield  {author} {\bibinfo {author} {\bibfnamefont {C.-H.}\ \bibnamefont
  {{Chuang}}}, \bibinfo {author} {\bibfnamefont {F.-S.}\ \bibnamefont
  {{Kitaura}}}, \bibinfo {author} {\bibfnamefont {Y.}~\bibnamefont {{Liang}}},
  \bibinfo {author} {\bibfnamefont {A.}~\bibnamefont {{Font-Ribera}}}, \bibinfo
  {author} {\bibfnamefont {C.}~\bibnamefont {{Zhao}}}, \bibinfo {author}
  {\bibfnamefont {P.}~\bibnamefont {{McDonald}}}, \ and\ \bibinfo {author}
  {\bibfnamefont {C.}~\bibnamefont {{Tao}}},\ }\bibfield  {title} {\enquote
  {\bibinfo {title} {{Linear redshift space distortions for cosmic voids based
  on galaxies in redshift space}},}\ }\href {\doibase
  10.1103/PhysRevD.95.063528} {\bibfield  {journal} {\bibinfo  {journal}
  {\prd}\ }\textbf {\bibinfo {volume} {95}},\ \bibinfo {eid} {063528} (\bibinfo
  {year} {2017})},\ \Eprint {http://arxiv.org/abs/1605.05352}
  {arXiv:1605.05352} \BibitemShut {NoStop}%
\bibitem [{\citenamefont {{Ata}}\ \emph {et~al.}(2017)\citenamefont {{Ata}},
  \citenamefont {{Kitaura}}, \citenamefont {{Chuang}}, \citenamefont
  {{Rodr{\'{\i}}guez-Torres}}, \citenamefont {{Angulo}}, \citenamefont
  {{Ferraro}}, \citenamefont {{Gil-Mar{\'{\i}}n}}, \citenamefont {{McDonald}},
  \citenamefont {{Hern{\'a}ndez Monteagudo}}, \citenamefont {{M{\"u}ller}},
  \citenamefont {{Yepes}}, \citenamefont {{Autefage}}, \citenamefont
  {{Baumgarten}}, \citenamefont {{Beutler}}, \citenamefont {{Brownstein}},
  \citenamefont {{Burden}}, \citenamefont {{Eisenstein}}, \citenamefont
  {{Guo}}, \citenamefont {{Ho}}, \citenamefont {{McBride}}, \citenamefont
  {{Neyrinck}}, \citenamefont {{Olmstead}}, \citenamefont {{Padmanabhan}},
  \citenamefont {{Percival}}, \citenamefont {{Prada}}, \citenamefont {{Rossi}},
  \citenamefont {{S{\'a}nchez}}, \citenamefont {{Schlegel}}, \citenamefont
  {{Schneider}}, \citenamefont {{Seo}}, \citenamefont {{Streblyanska}},
  \citenamefont {{Tinker}}, \citenamefont {{Tojeiro}},\ and\ \citenamefont
  {{Vargas-Magana}}}]{2017MNRAS.467.3993A}%
  \BibitemOpen
  \bibfield  {author} {\bibinfo {author} {\bibfnamefont {M.}~\bibnamefont
  {{Ata}}}, \bibinfo {author} {\bibfnamefont {F.-S.}\ \bibnamefont
  {{Kitaura}}}, \bibinfo {author} {\bibfnamefont {C.-H.}\ \bibnamefont
  {{Chuang}}}, \bibinfo {author} {\bibfnamefont {S.}~\bibnamefont
  {{Rodr{\'{\i}}guez-Torres}}}, \bibinfo {author} {\bibfnamefont {R.~E.}\
  \bibnamefont {{Angulo}}}, \bibinfo {author} {\bibfnamefont {S.}~\bibnamefont
  {{Ferraro}}}, \bibinfo {author} {\bibfnamefont {H.}~\bibnamefont
  {{Gil-Mar{\'{\i}}n}}}, \bibinfo {author} {\bibfnamefont {P.}~\bibnamefont
  {{McDonald}}}, \bibinfo {author} {\bibfnamefont {C.}~\bibnamefont
  {{Hern{\'a}ndez Monteagudo}}}, \bibinfo {author} {\bibfnamefont
  {V.}~\bibnamefont {{M{\"u}ller}}}, \bibinfo {author} {\bibfnamefont
  {G.}~\bibnamefont {{Yepes}}}, \bibinfo {author} {\bibfnamefont
  {M.}~\bibnamefont {{Autefage}}}, \bibinfo {author} {\bibfnamefont
  {F.}~\bibnamefont {{Baumgarten}}}, \bibinfo {author} {\bibfnamefont
  {F.}~\bibnamefont {{Beutler}}}, \bibinfo {author} {\bibfnamefont {J.~R.}\
  \bibnamefont {{Brownstein}}}, \bibinfo {author} {\bibfnamefont
  {A.}~\bibnamefont {{Burden}}}, \bibinfo {author} {\bibfnamefont {D.~J.}\
  \bibnamefont {{Eisenstein}}}, \bibinfo {author} {\bibfnamefont
  {H.}~\bibnamefont {{Guo}}}, \bibinfo {author} {\bibfnamefont
  {S.}~\bibnamefont {{Ho}}}, \bibinfo {author} {\bibfnamefont {C.}~\bibnamefont
  {{McBride}}}, \bibinfo {author} {\bibfnamefont {M.}~\bibnamefont
  {{Neyrinck}}}, \bibinfo {author} {\bibfnamefont {M.~D.}\ \bibnamefont
  {{Olmstead}}}, \bibinfo {author} {\bibfnamefont {N.}~\bibnamefont
  {{Padmanabhan}}}, \bibinfo {author} {\bibfnamefont {W.~J.}\ \bibnamefont
  {{Percival}}}, \bibinfo {author} {\bibfnamefont {F.}~\bibnamefont {{Prada}}},
  \bibinfo {author} {\bibfnamefont {G.}~\bibnamefont {{Rossi}}}, \bibinfo
  {author} {\bibfnamefont {A.~G.}\ \bibnamefont {{S{\'a}nchez}}}, \bibinfo
  {author} {\bibfnamefont {D.}~\bibnamefont {{Schlegel}}}, \bibinfo {author}
  {\bibfnamefont {D.~P.}\ \bibnamefont {{Schneider}}}, \bibinfo {author}
  {\bibfnamefont {H.-J.}\ \bibnamefont {{Seo}}}, \bibinfo {author}
  {\bibfnamefont {A.}~\bibnamefont {{Streblyanska}}}, \bibinfo {author}
  {\bibfnamefont {J.}~\bibnamefont {{Tinker}}}, \bibinfo {author}
  {\bibfnamefont {R.}~\bibnamefont {{Tojeiro}}}, \ and\ \bibinfo {author}
  {\bibfnamefont {M.}~\bibnamefont {{Vargas-Magana}}},\ }\bibfield  {title}
  {\enquote {\bibinfo {title} {{The clustering of galaxies in the completed
  SDSS-III Baryon Oscillation Spectroscopic Survey: cosmic flows and cosmic web
  from luminous red galaxies}},}\ }\href {\doibase 10.1093/mnras/stx178}
  {\bibfield  {journal} {\bibinfo  {journal} {\mnras}\ }\textbf {\bibinfo
  {volume} {467}},\ \bibinfo {pages} {3993--4014} (\bibinfo {year} {2017})},\
  \Eprint {http://arxiv.org/abs/1605.09745} {arXiv:1605.09745} \BibitemShut
  {NoStop}%
\bibitem [{\citenamefont {{En{\ss}lin}}\ \emph {et~al.}(2009)\citenamefont
  {{En{\ss}lin}}, \citenamefont {{Frommert}},\ and\ \citenamefont
  {{Kitaura}}}]{2009PhRvD..80j5005E}%
  \BibitemOpen
  \bibfield  {author} {\bibinfo {author} {\bibfnamefont {T.~A.}\ \bibnamefont
  {{En{\ss}lin}}}, \bibinfo {author} {\bibfnamefont {M.}~\bibnamefont
  {{Frommert}}}, \ and\ \bibinfo {author} {\bibfnamefont {F.~S.}\ \bibnamefont
  {{Kitaura}}},\ }\bibfield  {title} {\enquote {\bibinfo {title} {{Information
  field theory for cosmological perturbation reconstruction and nonlinear
  signal analysis}},}\ }\href {\doibase 10.1103/PhysRevD.80.105005} {\bibfield
  {journal} {\bibinfo  {journal} {\prd}\ }\textbf {\bibinfo {volume} {80}},\
  \bibinfo {eid} {105005} (\bibinfo {year} {2009})},\ \Eprint
  {http://arxiv.org/abs/0806.3474} {arXiv:0806.3474} \BibitemShut {NoStop}%
\bibitem [{\citenamefont {{Di Dio}}\ \emph {et~al.}(2017)\citenamefont {{Di
  Dio}}, \citenamefont {{Perrier}}, \citenamefont {{Durrer}}, \citenamefont
  {{Marozzi}}, \citenamefont {{Moradinezhad Dizgah}}, \citenamefont
  {{Nore{\~n}a}},\ and\ \citenamefont {{Riotto}}}]{2017JCAP...03..006D}%
  \BibitemOpen
  \bibfield  {author} {\bibinfo {author} {\bibfnamefont {E.}~\bibnamefont {{Di
  Dio}}}, \bibinfo {author} {\bibfnamefont {H.}~\bibnamefont {{Perrier}}},
  \bibinfo {author} {\bibfnamefont {R.}~\bibnamefont {{Durrer}}}, \bibinfo
  {author} {\bibfnamefont {G.}~\bibnamefont {{Marozzi}}}, \bibinfo {author}
  {\bibfnamefont {A.}~\bibnamefont {{Moradinezhad Dizgah}}}, \bibinfo {author}
  {\bibfnamefont {J.}~\bibnamefont {{Nore{\~n}a}}}, \ and\ \bibinfo {author}
  {\bibfnamefont {A.}~\bibnamefont {{Riotto}}},\ }\bibfield  {title} {\enquote
  {\bibinfo {title} {{Non-Gaussianities due to relativistic corrections to the
  observed galaxy bispectrum}},}\ }\href {\doibase
  10.1088/1475-7516/2017/03/006} {\bibfield  {journal} {\bibinfo  {journal}
  {Journal of Cosmology and Astro-Particle Physics}\ }\textbf {\bibinfo
  {volume} {2017}},\ \bibinfo {eid} {006} (\bibinfo {year} {2017})}\BibitemShut
  {NoStop}%
\bibitem [{\citenamefont {{Font-Ribera}}\ \emph {et~al.}(2014)\citenamefont
  {{Font-Ribera}}, \citenamefont {{McDonald}}, \citenamefont {{Mostek}},
  \citenamefont {{Reid}}, \citenamefont {{Seo}},\ and\ \citenamefont
  {{Slosar}}}]{2014JCAP...05..023F}%
  \BibitemOpen
  \bibfield  {author} {\bibinfo {author} {\bibfnamefont {A.}~\bibnamefont
  {{Font-Ribera}}}, \bibinfo {author} {\bibfnamefont {P.}~\bibnamefont
  {{McDonald}}}, \bibinfo {author} {\bibfnamefont {N.}~\bibnamefont
  {{Mostek}}}, \bibinfo {author} {\bibfnamefont {B.~A.}\ \bibnamefont
  {{Reid}}}, \bibinfo {author} {\bibfnamefont {H.-J.}\ \bibnamefont {{Seo}}}, \
  and\ \bibinfo {author} {\bibfnamefont {A.}~\bibnamefont {{Slosar}}},\
  }\bibfield  {title} {\enquote {\bibinfo {title} {{DESI and other Dark Energy
  experiments in the era of neutrino mass measurements}},}\ }\href {\doibase
  10.1088/1475-7516/2014/05/023} {\bibfield  {journal} {\bibinfo  {journal}
  {\jcap}\ }\textbf {\bibinfo {volume} {5}},\ \bibinfo {eid} {023} (\bibinfo
  {year} {2014})},\ \Eprint {http://arxiv.org/abs/1308.4164} {arXiv:1308.4164}
  \BibitemShut {NoStop}%
\bibitem [{\citenamefont {{DESI Collaboration}}\ \emph
  {et~al.}(2016)\citenamefont {{DESI Collaboration}}, \citenamefont
  {{Aghamousa}}, \citenamefont {{Aguilar}}, \citenamefont {{Ahlen}},
  \citenamefont {{Alam}}, \citenamefont {{Allen}}, \citenamefont {{Allende
  Prieto}}, \citenamefont {{Annis}}, \citenamefont {{Bailey}}, \citenamefont
  {{Balland}},\ and\ \citenamefont {et~al.}}]{2016arXiv161100036D}%
  \BibitemOpen
  \bibfield  {author} {\bibinfo {author} {\bibnamefont {{DESI Collaboration}}},
  \bibinfo {author} {\bibfnamefont {A.}~\bibnamefont {{Aghamousa}}}, \bibinfo
  {author} {\bibfnamefont {J.}~\bibnamefont {{Aguilar}}}, \bibinfo {author}
  {\bibfnamefont {S.}~\bibnamefont {{Ahlen}}}, \bibinfo {author} {\bibfnamefont
  {S.}~\bibnamefont {{Alam}}}, \bibinfo {author} {\bibfnamefont {L.~E.}\
  \bibnamefont {{Allen}}}, \bibinfo {author} {\bibfnamefont {C.}~\bibnamefont
  {{Allende Prieto}}}, \bibinfo {author} {\bibfnamefont {J.}~\bibnamefont
  {{Annis}}}, \bibinfo {author} {\bibfnamefont {S.}~\bibnamefont {{Bailey}}},
  \bibinfo {author} {\bibfnamefont {C.}~\bibnamefont {{Balland}}}, \ and\
  \bibinfo {author} {\bibnamefont {et~al.}},\ }\bibfield  {title} {\enquote
  {\bibinfo {title} {{The DESI Experiment Part I: Science,Targeting, and Survey
  Design}},}\ }\href@noop {} {\bibfield  {journal} {\bibinfo  {journal} {ArXiv
  e-prints}\ } (\bibinfo {year} {2016})},\ \Eprint
  {http://arxiv.org/abs/1611.00036} {arXiv:1611.00036 [astro-ph.IM]}
  \BibitemShut {NoStop}%
\bibitem [{\citenamefont {{Matarrese}}\ and\ \citenamefont
  {{Pietroni}}(2007)}]{2007JCAP...06..026M}%
  \BibitemOpen
  \bibfield  {author} {\bibinfo {author} {\bibfnamefont {S.}~\bibnamefont
  {{Matarrese}}}\ and\ \bibinfo {author} {\bibfnamefont {M.}~\bibnamefont
  {{Pietroni}}},\ }\bibfield  {title} {\enquote {\bibinfo {title} {{Resumming
  cosmic perturbations}},}\ }\href {\doibase 10.1088/1475-7516/2007/06/026}
  {\bibfield  {journal} {\bibinfo  {journal} {Journal of Cosmology and
  Astro-Particle Physics}\ }\textbf {\bibinfo {volume} {6}},\ \bibinfo {pages}
  {26--+} (\bibinfo {year} {2007})},\ \Eprint
  {http://arxiv.org/abs/arXiv:astro-ph/0703563} {arXiv:astro-ph/0703563}
  \BibitemShut {NoStop}%
\bibitem [{\citenamefont {{Izumi}}\ and\ \citenamefont
  {{Soda}}(2007)}]{2007PhRvD..76h3517I}%
  \BibitemOpen
  \bibfield  {author} {\bibinfo {author} {\bibfnamefont {K.}~\bibnamefont
  {{Izumi}}}\ and\ \bibinfo {author} {\bibfnamefont {J.}~\bibnamefont
  {{Soda}}},\ }\bibfield  {title} {\enquote {\bibinfo {title} {{Renormalized
  Newtonian cosmic evolution with primordial non-Gaussianity}},}\ }\href
  {\doibase 10.1103/PhysRevD.76.083517} {\bibfield  {journal} {\bibinfo
  {journal} {\prd}\ }\textbf {\bibinfo {volume} {76}},\ \bibinfo {pages}
  {083517--+} (\bibinfo {year} {2007})},\ \Eprint
  {http://arxiv.org/abs/arXiv:0706.1604} {arXiv:0706.1604} \BibitemShut
  {NoStop}%
\bibitem [{\citenamefont {{Rosten}}(2008)}]{2008JCAP...01..029R}%
  \BibitemOpen
  \bibfield  {author} {\bibinfo {author} {\bibfnamefont {O.~J.}\ \bibnamefont
  {{Rosten}}},\ }\bibfield  {title} {\enquote {\bibinfo {title} {{A comment on
  the path integral approach to cosmological perturbation theory}},}\ }\href
  {\doibase 10.1088/1475-7516/2008/01/029} {\bibfield  {journal} {\bibinfo
  {journal} {\jcap}\ }\textbf {\bibinfo {volume} {1}},\ \bibinfo {eid} {029}
  (\bibinfo {year} {2008})},\ \Eprint {http://arxiv.org/abs/0711.0867}
  {arXiv:0711.0867} \BibitemShut {NoStop}%
\bibitem [{\citenamefont {{Matarrese}}\ and\ \citenamefont
  {{Pietroni}}(2008)}]{2008MPLA...23...25M}%
  \BibitemOpen
  \bibfield  {author} {\bibinfo {author} {\bibfnamefont {S.}~\bibnamefont
  {{Matarrese}}}\ and\ \bibinfo {author} {\bibfnamefont {M.}~\bibnamefont
  {{Pietroni}}},\ }\bibfield  {title} {\enquote {\bibinfo {title} {{Baryonic
  Acoustic Oscillations via the Renormalization Group}},}\ }\href {\doibase
  10.1142/S0217732308026182} {\bibfield  {journal} {\bibinfo  {journal} {Modern
  Physics Letters A}\ }\textbf {\bibinfo {volume} {23}},\ \bibinfo {pages}
  {25--32} (\bibinfo {year} {2008})},\ \Eprint
  {http://arxiv.org/abs/arXiv:astro-ph/0702653} {arXiv:astro-ph/0702653}
  \BibitemShut {NoStop}%
\bibitem [{\citenamefont {{Floerchinger}}\ \emph {et~al.}(2017)\citenamefont
  {{Floerchinger}}, \citenamefont {{Garny}}, \citenamefont {{Tetradis}},\ and\
  \citenamefont {{Wiedemann}}}]{2017JCAP...01..048F}%
  \BibitemOpen
  \bibfield  {author} {\bibinfo {author} {\bibfnamefont {S.}~\bibnamefont
  {{Floerchinger}}}, \bibinfo {author} {\bibfnamefont {M.}~\bibnamefont
  {{Garny}}}, \bibinfo {author} {\bibfnamefont {N.}~\bibnamefont {{Tetradis}}},
  \ and\ \bibinfo {author} {\bibfnamefont {U.~A.}\ \bibnamefont
  {{Wiedemann}}},\ }\bibfield  {title} {\enquote {\bibinfo {title}
  {{Renormalization-group flow of the effective action of cosmological
  large-scale structures}},}\ }\href {\doibase 10.1088/1475-7516/2017/01/048}
  {\bibfield  {journal} {\bibinfo  {journal} {\jcap}\ }\textbf {\bibinfo
  {volume} {1}},\ \bibinfo {eid} {048} (\bibinfo {year} {2017})},\ \Eprint
  {http://arxiv.org/abs/1607.03453} {arXiv:1607.03453} \BibitemShut {NoStop}%
\bibitem [{\citenamefont {{En{\ss}lin}}(2018)}]{2018arXiv180403350E}%
  \BibitemOpen
  \bibfield  {author} {\bibinfo {author} {\bibfnamefont {Torsten~A.}\
  \bibnamefont {{En{\ss}lin}}},\ }\bibfield  {title} {\enquote {\bibinfo
  {title} {{Information theory for fields}},}\ }\href@noop {} {\bibfield
  {journal} {\bibinfo  {journal} {ArXiv e-prints}\ ,\ \bibinfo {eid}
  {arXiv:1804.03350}} (\bibinfo {year} {2018})},\ \Eprint
  {http://arxiv.org/abs/1804.03350} {arXiv:1804.03350 [astro-ph.CO]}
  \BibitemShut {NoStop}%
\end{thebibliography}%

\end{document}